




\font\twelverm=cmr10 scaled 1200   \font\twelvei=cmmi10 scaled 1200
\font\twelvesy=cmsy10 scaled 1200  \font\twelveex=cmex10 scaled 1200
\font\twelvebf=cmbx10 scaled 1200  \font\twelvesl=cmsl10 scaled 1200
\font\twelvett=cmtt10 scaled 1200  \font\twelveit=cmti10 scaled 1200
\font\twelvesc=cmcsc10 scaled 1200
\skewchar\twelvei='177   \skewchar\twelvesy='60


\def\twelvepoint{\normalbaselineskip=12.4pt plus 0.1pt minus 0.1pt
  \abovedisplayskip 12.4pt plus 3pt minus 9pt
  \belowdisplayskip 12.4pt plus 3pt minus 9pt
  \abovedisplayshortskip 0pt plus 3pt
  \belowdisplayshortskip 7.2pt plus 3pt minus 4pt
  \smallskipamount=3.6pt plus1.2pt minus1.2pt
  \medskipamount=7.2pt plus2.4pt minus2.4pt
  \bigskipamount=14.4pt plus4.8pt minus4.8pt
  \def\rm{\fam0\twelverm}          \def\it{\fam\itfam\twelveit}%
  \def\sl{\fam\slfam\twelvesl}     \def\bf{\fam\bffam\twelvebf}%
  \def\mit{\fam 1}                 \def\cal{\fam 2}%
  \def\sc{\twelvesc}               \def\tt{\twelvett}
  \def\sf{\twelvesf}
  \textfont0=\twelverm \scriptfont0=\tenrm
	\scriptscriptfont0=\sevenrm
  \textfont1=\twelvei  \scriptfont1=\teni
	\scriptscriptfont1=\seveni
  \textfont2=\twelvesy \scriptfont2=\tensy
	\scriptscriptfont2=\sevensy
  \textfont3=\twelveex \scriptfont3=\twelveex
	\scriptscriptfont3=\twelveex
  \textfont\itfam=\twelveit
  \textfont\slfam=\twelvesl
  \textfont\bffam=\twelvebf \scriptfont\bffam=\tenbf
  \scriptscriptfont\bffam=\sevenbf
  \normalbaselines\rm}


\def\beginlinemode{\endmode
  \begingroup\parskip=0pt \obeylines\def\\{\par}
	\def\endmode{\par\endgroup}}

\def\beginparmode{\endmode
  \begingroup \def\endmode{\par\endgroup}}
\let\endmode=\par
{\obeylines\gdef\
{}}

\newcount\firstpageno
\firstpageno=2
\footline={\ifnum\pageno<\firstpageno{\hfil}\else{\hfil\twelverm
	\folio\hfil}\fi}

\def\raggedcenter{\leftskip=4em plus 12em \rightskip=\leftskip
  \parindent=0pt \parfillskip=0pt \spaceskip=.3333em
  \xspaceskip=.5em \pretolerance=9999 \tolerance=9999
  \hyphenpenalty=9999 \exhyphenpenalty=9999 }

\hsize=6.5truein
\hoffset=0pt
\vsize=8.9truein
\voffset=0pt
\parskip=\medskipamount

\twelvepoint            

\overfullrule=0pt       

\def\head#1{
  \goodbreak\vskip 0.5truein
  {\immediate\write16{#1}
   \raggedcenter \uppercase{#1}\par}
   \nobreak\vskip 0.25truein\nobreak}

\def\subhead#1{
  \vskip 0.25truein
  {\raggedcenter {#1} \par}
   \nobreak\vskip 0.25truein\nobreak}


\def\refto#1{$^{#1}$}

\def\references
  {\head{References}
   \beginparmode
   \frenchspacing \parindent=0pt \leftskip=1truecm
   \parskip=8pt plus 3pt \everypar{\hangindent=\parindent}}

\gdef\refis#1{\item{#1.\ }}

\def\endreferences{\beginparmode}

\def\ref#1{Ref.~#1}
\def\Ref#1{Ref.~#1}
\def\[#1]{[\cite{#1}]}
\def\cite#1{{#1}}
\def\(#1){(\call{#1})}
\def\call#1{{#1}}


\catcode`@=11
\newcount\tagnumber\tagnumber=0

\immediate\newwrite\eqnfile
\newif\if@qnfile\@qnfilefalse
\def\write@qn#1{}
\def\writenew@qn#1{}
\def\w@rnwrite#1{\write@qn{#1}\message{#1}}
\def\@rrwrite#1{\write@qn{#1}\errmessage{#1}}

\def\taghead#1{\gdef\t@ghead{#1}\global\tagnumber=0}
\def\t@ghead{}

\expandafter\def\csname @qnnum-3\endcsname
  {{\t@ghead\advance\tagnumber by -3\relax\number\tagnumber}}
\expandafter\def\csname @qnnum-2\endcsname
  {{\t@ghead\advance\tagnumber by -2\relax\number\tagnumber}}
\expandafter\def\csname @qnnum-1\endcsname
  {{\t@ghead\advance\tagnumber by -1\relax\number\tagnumber}}
\expandafter\def\csname @qnnum0\endcsname
  {\t@ghead\number\tagnumber}
\expandafter\def\csname @qnnum+1\endcsname
  {{\t@ghead\advance\tagnumber by 1\relax\number\tagnumber}}
\expandafter\def\csname @qnnum+2\endcsname
  {{\t@ghead\advance\tagnumber by 2\relax\number\tagnumber}}
\expandafter\def\csname @qnnum+3\endcsname
  {{\t@ghead\advance\tagnumber by 3\relax\number\tagnumber}}

\def\equationfile{%
  \@qnfiletrue\immediate\openout\eqnfile=\jobname.eqn%
  \def\write@qn##1{\if@qnfile\immediate\write\eqnfile{##1}\fi}
  \def\writenew@qn##1{\if@qnfile\immediate\write\eqnfile
    {\noexpand\tag{##1} = (\t@ghead\number\tagnumber)}\fi}
}

\def\callall#1{\xdef#1##1{#1{\noexpand\call{##1}}}}
\def\call#1{\each@rg\callr@nge{#1}}

\def\each@rg#1#2{{\let\thecsname=#1\expandafter\first@rg#2,\end,}}
\def\first@rg#1,{\thecsname{#1}\apply@rg}
\def\apply@rg#1,{\ifx\end#1\let\next=\relax%
\else,\thecsname{#1}\let\next=\apply@rg\fi\next}

\def\callr@nge#1{\calldor@nge#1-\end-}
\def\callr@ngeat#1\end-{#1}
\def\calldor@nge#1-#2-{\ifx\end#2\@qneatspace#1 %
  \else\calll@@p{#1}{#2}\callr@ngeat\fi}
\def\calll@@p#1#2{\ifnum#1>#2{\
	@rrwrite{Equation range #1-#2\space is bad.}
\errhelp{If you call a series of equations
 by the notation M-N, then M and N must be integers, and N
 must be greater than or equal to M.}}\else%
 {\count0=#1\count1=#2\advance\count1 by1\relax
  \expandafter\@qncall\the\count0,%
  \loop\advance\count0 by1\relax%
    \ifnum\count0<\count1,\expandafter\@qncall\the\count0,%
  \repeat}\fi}

\def\@qneatspace#1#2 {\@qncall#1#2,}
\def\@qncall#1,{\ifunc@lled{#1}{\def\next{#1}\ifx\next\empty\else
  \w@rnwrite{Equation number \noexpand\(>>#1<<)
  has not been defined yet.}
  >>#1<<\fi}\else\csname @qnnum#1\endcsname\fi}

\let\eqnono=\eqno
\def\eqno(#1){\tag#1}
\def\tag#1$${\eqnono(\displayt@g#1 )$$}

\def\aligntag#1\endaligntag
  $${\gdef\tag##1\\{&(##1 )\cr}\eqalignno{#1\\}$$
  \gdef\tag##1$${\eqnono(\displayt@g##1 )$$}}

\def\eqalignno#1{\displ@y \tabskip\centering
  \halign to\displaywidth{\hfil$\displaystyle{##}$\tabskip\z@skip
    &$\displaystyle{{}##}$\hfil\tabskip\centering
    &\llap{$\displayt@gpar##$}\tabskip\z@skip\crcr
    #1\crcr}}

\def\displayt@gpar(#1){(\displayt@g#1 )}

\def\displayt@g#1 {\rm\ifunc@lled{#1}\global\advance\tagnumber by1
      {\def\next{#1}\ifx\next\empty\else\expandafter
      \xdef\csname @qnnum#1\endcsname{\t@ghead\number\tagnumber}\fi}
  \writenew@qn{#1}\t@ghead\number\tagnumber\else
    {\edef\next{\t@ghead\number\tagnumber}%
    \expandafter\ifx\csname @qnnum#1\endcsname\next\else
    \w@rnwrite{Equation \noexpand\tag{#1} is a duplicate number.}\fi}
  \csname @qnnum#1\endcsname\fi}

\def\ifunc@lled#1{\expandafter\ifx\csname @qnnum#1\endcsname\relax}

\let\@qnend=\end\gdef\end{\if@qnfile
\immediate\write16{Equation numbers written on
  []\jobname.EQN.}\fi\@qnend}

\catcode`@=12


\catcode`@=11
\newcount\r@fcount \r@fcount=0
\newcount\r@fcurr
\immediate\newwrite\reffile
\newif\ifr@ffile\r@ffilefalse
\def\w@rnwrite#1{\ifr@ffile\immediate\write\reffile{#1}
  \fi\message{#1}}

\def\writer@f#1>>{}
\def\referencefile{
  \r@ffiletrue\immediate\openout\reffile=\jobname.ref%
  \def\writer@f##1>>{\ifr@ffile\immediate\write\reffile%
    {\noexpand\refis{##1} = \csname r@fnum##1\endcsname = %
     \expandafter\expandafter\expandafter\strip@t\expandafter%
     \meaning\csname r@ftext\csname r@fnum##1\endcsname
     \endcsname}\fi}%
  \def\strip@t##1>>{}}

\def\citeall#1{\xdef#1##1{#1{\noexpand\cite{##1}}}}
\def\cite#1{\each@rg\citer@nge{#1}}

\def\each@rg#1#2{{\let\thecsname=#1\expandafter\first@rg#2,\end,}}
\def\first@rg#1,{\thecsname{#1}\apply@rg}
\def\apply@rg#1,{\ifx\end#1\let\next=\relax%
\else,\thecsname{#1}\let\next=\apply@rg\fi\next}%

\def\citer@nge#1{\citedor@nge#1-\end-}
\def\citer@ngeat#1\end-{#1}
\def\citedor@nge#1-#2-{\ifx\end#2\r@featspace#1
  \else\citel@@p{#1}{#2}\citer@ngeat\fi}
\def\citel@@p#1#2{\ifnum#1>#2{
	\errmessage{Reference range #1-#2\space is bad.}%
    \errhelp{If you cite a series of references
    by the notation M-N, then M and  N must be integers,
    and N must be greater than or equal to M.}}\else%
 {\count0=#1\count1=#2\advance\count1 by1\relax
  \expandafter\r@fcite\the\count0,%
  \loop\advance\count0 by1\relax
    \ifnum\count0<\count1,\expandafter\r@fcite\the\count0,%
  \repeat}\fi}

\def\r@featspace#1#2 {\r@fcite#1#2,}
\def\r@fcite#1,{\ifuncit@d{#1}%
    \newr@f{#1}%
    \expandafter\gdef\csname r@ftext\number\r@fcount\endcsname%
                     {\message{Reference #1 to be supplied.}%
                      \writer@f#1>>#1 to be supplied.\par}%
 \fi%
 \csname r@fnum#1\endcsname}
\def\ifuncit@d#1{\expandafter\ifx\csname r@fnum#1\endcsname\relax}%
\def\newr@f#1{\global\advance\r@fcount by1%
    \expandafter\xdef\csname r@fnum#1\endcsname{\number\r@fcount}}

\let\r@fis=\refis
\def\refis#1#2#3\par{\ifuncit@d{#1}%
   \newr@f{#1}%
   \w@rnwrite{Reference #1=\number
   \r@fcount\space is not cited up to now.}\fi%
  \expandafter\gdef\csname r@ftext\csname r@fnum#1\endcsname
  \endcsname%
  {\writer@f#1>>#2#3\par}}

\def\ignoreuncited{
   \def\refis##1##2##3\par{\ifuncit@d{##1}%
     \else\expandafter\gdef\csname r@ftext\csname r@fnum##1
   \endcsname\endcsname%
     {\writer@f##1>>##2##3\par}\fi}}

\def\r@ferr{\endreferences\errmessage{I was expecting to see
\noexpand\endreferences before now;  I have inserted it here.}}
\let\r@ferences=\references
\def\references{\r@ferences\def\endmode{\r@ferr\par\endgroup}}

\let\endr@ferences=\endreferences
\def\endreferences{\r@fcurr=0%
  {\loop\ifnum\r@fcurr<\r@fcount%
    \advance\r@fcurr by 1\relax\expandafter\r@fis
    \expandafter{\number\r@fcurr}%
    \csname r@ftext\number\r@fcurr\endcsname%
  \repeat}\gdef\r@ferr{}\endr@ferences}


\let\r@fend=\endpaper\gdef\endpaper{\ifr@ffile
\immediate\write16{Cross References written on
  []\jobname.REF.}\fi\r@fend}

\catcode`@=12

\citeall\refto		
\citeall\ref		%
\citeall\Ref		%



\def\undertext#1{$\underline{\hbox{#1}}$}
\def\frac#1#2{{#1 \over #2}}
\def\12{{1\over2}}
\def\square{\kern1pt\vbox{\hrule height 1.2pt
\hbox{\vrule width 1.2pt\hskip 3pt
   \vbox{\vskip 6pt}\hskip 3pt
   \vrule width 0.6pt}\hrule height 0.6pt}\kern1pt}


\baselineskip=\normalbaselineskip \multiply\baselineskip by 2

\null\vskip 3pt plus 0.2fill
   \beginlinemode \baselineskip=\normalbaselineskip
\multiply\baselineskip by 2 \raggedcenter \bf

Generalized Sums over Histories for Quantum Gravity
II. Simplicial Conifolds

\vskip 3pt plus 0.2fill \beginlinemode
  \baselineskip=\normalbaselineskip \raggedcenter\sc
        Kristin Schleich

\vskip 3pt plus 0.2fill \beginlinemode
  \baselineskip=\normalbaselineskip \raggedcenter\sc
        Donald M. Witt

\vskip 3pt plus 0.1fill \beginlinemode
   \baselineskip=\normalbaselineskip
  \multiply\baselineskip by 3 \divide\baselineskip by 2
\raggedcenter \sl

Department of Physics
University of British Columbia
Vancouver, British Columbia V6T 1Z1

\vskip 3pt plus 0.3fill \beginparmode
\baselineskip=\normalbaselineskip
  \multiply\baselineskip by 3 \divide\baselineskip by 2
\noindent ABSTRACT:
 This paper examines the issues involved with concretely implementing a
sum over conifolds in the formulation of Euclidean sums over histories
for gravity. The first step in precisely formulating any sum over
topological spaces is that  one  must have an algorithmically
implementable method of generating a list of all spaces in the set to
be summed over. This requirement causes well known problems in the
formulation of sums over  manifolds in four or more dimensions; there
is no algorithmic method of determining whether or not a topological
space is an n-manifold in five or more dimensions and the issue of
whether or not such an algorithm exists is open in four.
However, as this paper shows, conifolds are algorithmically decidable
in four dimensions.  Thus the set of 4-conifolds provides a starting
point for a concrete implementation of Euclidean sums over histories
in four dimensions.  Explicit algorithms for summing over various sets
of 4-conifolds are presented in the context of Regge calculus.

\vfill\eject
\beginparmode

\head{Introduction}

The sum over histories approach to formulating quantum amplitudes
provides a convenient and powerful tool for the study of many aspects
of quantum field theories. This approach is especially useful in the
study of topology and topology change in Euclidean gravitational
amplitudes; Euclidean gravitational histories consist both of a
topological space and a metric and thus a sum over histories
formulation  incorporates contributions from histories of different
topology in a very natural manner. For example, quantities such as the
transition amplitude between a set of manifolds $\Sigma^{n-1}$ with
metrics $h$,
$$\eqalignno{&G[\Sigma^{n-1}, h] = \sum_{(K^n,g)} \  \exp(-I[g])\cr
I[g] &=- \frac 1{16\pi G} \int_{K^n}  (R-2\Lambda) d\mu(g) - \frac
1{8\pi G}\int_{\Sigma^{n-1}}  K d\mu(h) &(topchange)\cr}$$
where $d\mu$ denotes the covariant volume element with respect to the
indicated metric are formed by taking the sum over an appropriate set
of compact physically distinct  $K^n$ and  $g$ weighted by the
Euclidean Einstein action.
Quantities formulated in terms of  sums over closed connected
topological spaces $K^n$ such as
$$<A> =\frac{\displaystyle \sum_{(K^n,g)} A(g)\exp\biggl(-I[g]
\biggr)}{\displaystyle \sum_{(K^n,g)} \exp\biggl(-I[g]
\biggr)}\eqno(partfn)$$
correspond to the expectation values of  geometrical objects $A$.  Thus
expressions such as \(topchange) and \(partfn) manifestly incorporate
contributions from histories of different topology.  They are
especially useful in the qualitative analysis of topology and topology
change because even without a precise implementation of the sum, they
can be evaluated in semiclassical approximation. These qualitative
studies have led to many very interesting results.\refto{genlist}
However, the study of topology change and other consequences of
topology in semiclassical approximation is limited as semiclassical
calculations consider only the contribution from a certain restricted
set of spaces $K^n$; those that are classical Euclidean instantons.
Therefore, though a useful qualitative guide, semiclassical evaluations
do not yield a method of studying topology in depth as they do not
include contributions from all topologies.

In order to go beyond such semiclassical evaluations,  it is necessary
to replace the heuristic expression \(topchange) with a more concrete
definition that explicitly
 implements the sum over histories with different topology. Of course
the goal of finding a complete, well defined sum over histories for
Euclidean gravity is  out of reach due to well known problems such as
the unboundedness of the Euclidean Einstein action and perturbative
nonrenormalizability. However, these problems are directly related to
properties of the quantum mechanics of the  metric alone. Moreover, the
topological aspects of forming a sum over histories can be isolated
from those involving the metric.  Indeed it is easy to observe that the
sum over histories for a quantum amplitude such as \(partfn) can be
written in the form of an explicit sum over spaces $K^n$ and a
functional integral over metrics $g$ on each topological space;
$$<A>=\frac{ \displaystyle \sum_{K^n}\int Dg A(g)\exp\biggl(-I[g]
\biggr)}{ \displaystyle \sum_{K^n}\int Dg \exp\biggl(-I[g] \biggr)}
\ \ .\eqno(fpartfn)$$
Thus the topological aspects of formulating a sum over histories can be
studied independently of those involving the metric. Moreover, these
topological aspects are of particular interest as they are not linked
to the dynamics of the metric itself. Therefore the issues that arise
in defining a sum over topological spaces such as manifolds or
conifolds should be relevant in any theory involving  such a sum, not
just Einstein gravity.

In attempting to concretely formulate any sum over topologies such as
that in \(fpartfn), two important and intimately related issues must be
addressed: what kinds of histories should be included in the sum and
whether or not  a sum over these spaces can be explicitly carried out.
The first  issue was addressed in part I  (Ref.[\cite{I}]).  The second
issue will be the topic of this  paper.

As discussed in part I, it is well known that the space of histories
for a theory is larger than the set of the classical histories of the
theory. For example, the space of histories for a field theory includes
nondifferentiable field configurations as well as smooth ones.
Similarly, one anticipates that the space of histories for Euclidean
gravity also includes some sort of histories that are less regular than
classical histories.  Now, a classical history in Euclidean gravity
consists of both a differentiable metric $g$  and smooth manifold
$M^n$.  It follows that a less regular history for gravity can be less
regular in two different ways; it can consist of a less regular metric
$g$, or a less regular topological space $K^n$. Additionally, these two
ways can be separated; nondifferentiable metrics can be defined on
smooth manifolds and conversely, regular metrics can be defined on
smooth topological spaces that are not manifolds.  Therefore, it is
natural to consider whether or not more general topological spaces than
manifolds should be included in a sum over topologies. Moreover, as the
topological generalizations are distinct from the metric
generalizations, this issue can be studied in the context of classical
paths. As discussed in detail in part I, semiclassical results indicate
that allowing only manifolds in the sum over histories is too
restrictive; limits of sequences of smooth manifolds lead to
non-manifold stationary points of the Euclidean action. Given that such
spaces occur in the semiclassical limit, it is logical that they should
be included in the space of histories for Euclidean functional
integrals for gravity. These non-manifold spaces are elements of a more
general set of topological spaces called conifolds:
\proclaim Definition {(1.1)}. A n-dimensional conifold $X^n$, $n\ge 2$,
is a metrizable space such that given any $x_0 \in  X^n$ there is an
open neighborhood $N_{x_0}$ and some closed connected (n-1)-manifold
$\Sigma_{x_0}^{n-1}$ such that $N_{x_0}$ is homeomorphic to the
interior of a cone over $\Sigma_{x_0}^{n-1}$ with $x_0$ mapped to the
apex of the cone.\par
\noindent Thus semiclassical results indicate that the topology of
histories in the space of histories for expressions such as \(fpartfn)
should be generalized to include smooth conifolds.

The second issue, how to explicitly carry out a sum over a set of
topological spaces such as manifolds or conifolds, will be discussed in
this paper.  Recall that the standard rule for formulating any sum over
histories is that only physically distinct histories should be included
in the sum. In  Euclidean gravity, two histories $(K^n,g)$ and
$({K'}^n, g')$ are physically distinct if they have metrics which are
not diffeomorphic; however, they are also physically distinct if their
underlying topological spaces $K^n $ and ${K'}^n$ are not
diffeomorphic. Thus, in formal terms, a sum over physically distinct
histories for Euclidean gravity
 should consist of a sum over smooth topological spaces that are not
diffeomorphic to each other. In four or more dimensions, it turns out
that topological spaces that are homeomorphic are not necessarily
diffeomorphic as they can admit different smooth structures. Thus a sum
over physically distinct topological spaces $K^n$ in \(fpartfn) must
include a sum over inequivalent smooth structures as well as a sum over
inequivalent (that is nonhomeomorphic) topological spaces.

Now, in order to make an expression such as \(fpartfn) concrete, one
must replace the heuristic summation sign with a concrete method of
taking such a sum. However, even though the number of closed
topological spaces  is countable, and even though the number of
distinct smooth structures on these spaces is countable, it turns out
that such a concrete method does not exist in all dimensions or for all
sets of topological spaces. For example, consider the formulation of a
sum over n-manifolds. One imagines explicitly implementing this sum by
first making a list of all physically distinct n-manifolds, that is a
list of all  manifolds that are not diffeomorphic to each other. One
then simply takes the sum in \(fpartfn) to be over  all distinct
n-manifolds in the list.  However, though such a technique sounds
reasonable, it turns out that it  cannot be carried out for general
dimension n.  There is no way to explicitly list all physically
distinct n-manifolds for $n\ge 4$ because whether or not two manifolds
are diffeomorphic cannot be determined by a finite procedure;
n-manifolds for $n\ge 4$ are not classifiable. Additionally,  there is
no known algorithm for classifying 3-manifolds.  Even worse, in five or
more dimensions, one can prove that there is no finite algorithm for
determining whether or not a given topological space satisfies the
definition of a manifold and there is no known finite algorithm for
doing so in four dimensions.  Moreover, without such a finite
algorithm, even the first step in concretely implementing an expression
of the form \(fpartfn)
cannot be carried out. Thus, as it stands, expressions such as
\(fpartfn) are not well defined for a sum over smooth n-manifolds in
arbitrary dimension.

It turns out that the ability to carry out an explicit formulation of a
sum over topological spaces depends intimately on two things; the set
of topological spaces at hand and the criteria by which they are to be
classified as distinct.  Different sets of topological spaces other
than the set of manifolds are  explicitly known to be algorithmically
decidable in four or more dimensions. Additionally
there are algorithmically decidable sets that include all n-manifolds
as a subset.  Thus by choosing a different set of topological spaces,
it is possible to explicitly construct a set of spaces that includes
all classical histories.

Given a constructible set of more general topological spaces, the next
task  is to find a set of unique representatives of physically distinct
topological spaces. As in the case of n-manifolds, the criteria for
doing so is  diffeomorphism invariance.  However, it turns out that the
problems with classifying n-manifolds for $n\ge 4$
 extends to any set of more general topological spaces that includes
all n-manifolds. Thus allowing more general sets of topological spaces
only addresses the first factor involved in the explicit construction
of sums over physically distinct histories, not the second.
 Therefore, in order to explicitly formulate a sum over distinct
topological spaces,
 the issue of when two topological spaces are to be identified as
distinct must be reexamined.

These observations about algorithmic decidability were used as
motivation by  Hartle for studying ``unruly topologies'' in
2-dimensional simplicial gravity.\refto{2dcase} Hartle argued
 that sums over 2-pseudomanifolds would produce the same qualitative
results in the classical limit as sums over 2-manifolds in expressions
such as \(fpartfn).  However, little concrete work has been done in
higher dimensions in either formulating sums over more general
topological spaces or exploring their consequences; there are many
algorithmically decidable spaces so without further information it is
difficult to select a viable candidate. However, part I of this paper
provides precisely the further information needed to select such a
candidate: a physically motivated set of spaces, conifolds.
Furthermore, it turns out that conifolds can be shown to be
algorithmically decidable in four or fewer dimensions. Thus, the set of
smooth conifolds provides a starting point for an explicit
implementation of the sum over topologies. Given this starting point,
different criteria for classifying conifolds can be formulated and
their consequences studied in terms of explicitly constructible
amplitudes.

This paper will give a comprehensive discussion of algorithmic
decidability and classifiability of both manifolds and conifolds in
general dimension and will provide  explicit implementations of sums
over conifolds in four dimensions.
In order to discuss the problems with implementing the sum over
topological spaces, it is necessary to have a finite representation of
the topological space. A well known method of doing so is to use
simplicial complexes.    The topology of the simplicial complex is
completely carried by the simple set of rules used to assemble it from
a countable set of elements, the simplices.  Section 2  will discuss
simplicial complexes and then present the definitions of combinatorial
manifolds and conifolds, the simplicial analogs of continuum manifolds
and conifolds. Section 3 will describe precisely how these simplicial
analogs are related to their smooth counterparts.  It turns out that in
less than seven dimensions, the set of smooth manifolds uniquely
corresponds to that of combinatorial manifolds and likewise for the
conifold case. Therefore combinatorial manifolds and conifolds are the
desired finite representations of smooth manifolds and conifolds in
less than seven dimensions. Additionally, a sum over combinatorial
spaces automatically incorporates a sum over smooth structures in less
than seven dimensions.  Section 4 will begin by discussing the
algorithmic decidability and classifiability of manifolds. Certain
important points about these results that are usually brushed over in
discussions of sums over manifolds in Euclidean gravity will be
emphasized; these points are especially relevant and
a misunderstanding of them can lead to erroneous claims. After this
discussion of the manifold case, the results on the algorithmic
decidability and classifiability of conifolds will be derived. Section
5 will discuss the consequences of the results of section 4 on the
definition of Euclidean functional integrals for gravity. It will
illustrate these consequences in terms of Regge calculus and provide
explicit algorithms for finding different  sets of distinct 4-conifolds
for implementing these sums. However, it is important to note that,
although Regge calculus is especially useful for studying effects of
topology and topology change numerically, the results on the explicit
formulation of sums over topology apply generally due to the results of
section 3.

\head{2.~Combinatorial Manifolds and Combinatorial Conifolds}

It is useful to begin by summarizing certain definitions and theorems
on simplicial complexes; even though simplicial complexes are well
known and often used in topology, there are many instances of different
authors using the same terminology  to refer to slightly different
objects.\refto{ST} Thus it is best to explicitly present the
definitions used for the reader's understanding. After this summary,
the definition of a combinatorial manifold and that of a combinatorial
conifold are provided. Finally, certain aspects of these definitions
particularly relevant to their use in this paper are discussed.

Simplicial complexes are a subset of the set of polyhedra discussed in
section 4 of part I. Thus it is useful to recall the definition of a
polyhedra.\refto{rs} First note that two subspaces $X$ and $Y$ of ${\bf
R}^n$ are said to be in {\it general position} if for any distinct
points $x_1,x_2\in X$ and $y_1,y_2\in Y$, the line segments connecting
$x_1$ to $y_1$ and $x_2$ to $y_2$ do not intersect. With this
terminology,
\proclaim Definition {(2.1)}.
Let two subspaces $X$ and $Y$ of ${\bf R}^n$ be in general position.
Their PL join is the union of all line segments joining points of $X$
to points of $Y$. The join will be denoted $XY$. \par
\noindent A PL cone is defined to be PL join of space $X$ with a point
$a$. It will be denoted by $aX$. Then, given the definition of a PL
cone,
\proclaim Definition {(2.2)}. A polyhedron $P$ is a subset of ${\bf
R}^n$ such that each point $p$ has a cone neighborhood $N=aL\subseteq
P$ where $L$ is a compact topological space. $L$ is called the link of
the neighborhood $N$.\par
\noindent The term polyhedra will be used in the strict sense of
Def.(2.2) in this paper
 in contrast to its usage in part I.  One can show that as topological
spaces, the PL join when it exists is homeomorphic to the topological
join $X*Y$. This means that as topological spaces, there is no
difference between the PL join and the topological join. The difference
 is that the PL join has more structure and can only be defined when
the two spaces can be positioned in this nice way.  In order to see
this, note that the natural equivalence of a polyhedra is given by
piecewise linear maps, referred to as PL maps.
\proclaim Definition {(2.3)}. A map $f:P \to Q$ between two polyhedra
is
piecewise linear if each point $a$ in $P$  has a cone neighborhood $N
= aL$ in $P$ such that $ f(\lambda a +\mu x) = \lambda f(a) + \mu f(x)$
where x is in $L$ and $\lambda, \mu \ge 0, \lambda + \mu =1$.

\noindent It is clear that a PL map from a join to itself cannot be
defined for a join that does not satisfy the conditions of Def.(2.1).
Thus the definition of a PL map characterizes the extra structure
inherent in the definition of a PL join.  Finally, of particular
interest are PL homeomorphisms, that is PL maps that are continuous and
have a continuous PL inverse. Two polyhedra are said to be PL
equivalent if there is a PL homeomorphism between them.

Given the previous definitions, the first step in defining a simplicial
complex is to define a simplex:\refto{spanier}

\proclaim Definition {(2.4)}. Let $v_1,v_2\ldots ,v_{n+1}$ be affinely
independent points\refto{points} in ${\bf R}^{n+1}$. An  n-simplex
$\sigma^n$ is the convex hull of these points:
$$ \sigma^n = \{x|
x=\sum_1^{n+1}\lambda_i v_i; \lambda_i \ge 0;\ \sum_1^{n+1} \lambda_i =
1 \}.$$

\noindent A 0-simplex is a point, a 1-simplex is a line segment or
edge, a 2-simplex is a triangle, a 3-simplex is a tetrahedra. Higher
dimensional simplices are generalizations of tetrahedra to higher
dimensions. A simplex spanned by a subset of the vertices is called a
{\it face}. Thus the vertices are all faces of a n-simplex; they are
0-simplices. Similarly, 1-simplices or edges formed from any two
distinct vertices are faces.  Note that by Def.(2.4), a simplex is
uniquely determined by its vertices. This property is very important
for both using and understanding simplicial complexes:

\proclaim Definition {(2.5)}. A  simplicial complex $K$ is a
topological space $|K|$ and a collection of simplices $K$ such that
\hfill\break
$\ \ $i) $|K|$ is a closed subset of some finite dimensional Euclidean
space \hfill\break
$\ \ $ ii) if $F$ is a face of a simplex $K$, then $F$ is also
contained in $K$ \hfill\break
$\ \ $ iii) if $B,C$ are simplices in $K$, then $B \cap C$ is a face of
both $B$ and $C$. \hfill\break
\noindent The topological space $|K|$ is the union of all simplices in
$K$.\par
A particularly simple example of a simplicial complex is provided by
the square in Figure 1a). It consists of four 0-simplices or vertices,
five 1-simplices and two 2-simplices. The rules for constructing the
square are given by enumerating which vertices   and edges are
contained in which triangle. This set of rules encodes the topology of
the square; these rules are conveniently given  the drawing itself,
i.e.  the two triangles have one edge in common. An alternate method
of giving these rules is to provide a list  of  which vertices are
contained in each simplex  of the complex as indicated in Figure 1a).
As higher dimensional simplices are uniquely specified by their
vertices, note that it is not really necessary to label them in any way
other than by the subset of the vertices they contain.

Observe that  simplicial complexes form a subset of the set of
polyhedra of Def.(2.2). However, some spaces that are homeomorphic to
polyhedra are not simplicial complexes because they do not satisfy all
the conditions of Def.(2.5). For example, a nice one dimensional
polyhedron is a circle represented by two line segments connected at
each endpoint in Figure 1c). This space is homeomorphic to a polyhedron
as it is a closed subset of ${\bf R}^2$; however it is not a simplicial
complex. Both line segments have the same endpoints $a,b$ and thus are
not uniquely determined by these endpoints.  This example may seem
obvious; however, it is easy to forget this important property of
simplicial complexes when dealing with higher dimensional spaces.
Finally, another simplicial description of the same topological space
as Figure 1a) is given in
Figure 1b) and some other examples of simplicial complexes are given
in Figures 2-6.

The dimension of a simplicial complex is the largest dimension of any
simplex contained in the complex. Condition {\sl i)}
implies that the simplicial complexes of Def.(2.5) are finite
dimensional. One can define infinite dimensional simplicial complexes
by changing  this condition.\refto{spanier} However, this paper is
concerned with finite dimensional spaces, so such a change is
unnecessary.  Similarly, abstract simplicial complexes can be defined
by using the last two conditions in  Def.(2.5) and replacing the first
condition with some topology for the space $|K|$ other than that
induced by Euclidean space  (See Appendix A). However, there is
generally no advantage to do this for finite dimensional complexes
because, if the topology used to define the complex is reasonable, then
the simplicial complex can be realized as a subset of Euclidean space.

Thus, a simplicial complex describes both the building blocks of the
space $|K|$ and gives the rules for how these building blocks are
connected together. Consequently the simplicial complex completely
describes the topology of the space $|K|$.  Finally, as each simplex in
$K$ is uniquely determined by its vertices, the simplicial complex
itself is uniquely determined the vertices and the rules for which
simplices they are contained in. It is clear that this property is
especially valuable for computational purposes.

Simplicial complexes can describe topological spaces containing
subspaces of different dimension, compact and noncompact spaces, and
spaces with boundary.  Thus, there are simplicial analogs of various
standard definitions in topology:
A {\it pure} simplicial complex is one in which every lower
dimensional simplex is contained in at least one n-simplex where n is
the  dimension of the simplicial complex. A {\it compact} simplicial
complex is one which contains a finite number of simplices. A {\it
connected } simplicial complex is one in which any two vertices are
connected by a sequence of edges.  One can easily verify that an
equivalent definition of a connected complex is that the underlying
space  $|K|$ of the complex is a connected topological space. The
properties of these definitions are illustrated by the examples of
Figure 2: All of the simplicial complexes drawn are compact. All of the
simplicial complexes are pure except Figure 2a); the flagpole is not
contained in any triangle. All of the simplicial complexes are
connected except Figure 2b). It is easy to see that these properties
can all be tested for simplicial complexes presented in the form of a
list as well these pictorial representations.

It is also interesting to define the simplicial analog of continuum
manifolds and conifolds with boundary.  It is clear that the notion of
boundary in the sense used in the continuum theory can only be made
meaningful for a subset of pure complexes. It turns out that there is
an  additional requirement needed to specify this subset:
\proclaim Definition {(2.6)}. A  nonbranching simplicial complex is a
simplicial complex of dimension n in which every (n-1)-simplex is
contained in at most two n-simplices.\par
\noindent For example, Figure 2c) is a
branching simplicial complex as it has an edge that is contained in
four triangles. Then
\proclaim Definition {(2.7)}. The boundary of a pure nonbranching
simplicial complex is the set of (n-1)-simplices that are faces of only
one n-simplex.\par
\noindent The boundary of fig.(1a)
is composed of the four
edges $\alpha, \beta, \gamma, \delta$ that are in only one triangle.
Figures 2b) and 2d) have no boundary by the above definition. Note
especially that the common vertex of the two tetrahedra of Figure 2d)
is not a boundary as a boundary is a (n-1) dimensional simplicial
complex by definition. This is completely in keeping with continuum
definition of boundary.\refto{pIdisc}

Def.(2.7) implies that a pure nonbranching complex without boundary is
one in which every (n-1)-simplex is contained in exactly two
n-simplices. Therefore, again there is an easily implementable test for
boundary in terms of any concrete description of the complex.  Finally,
it can be proven that the boundary of a pure nonbranching simplicial
complex is a topological invariant of the underlying space;  it is the
nonbranching condition that provides this necessary and desirable
property. Therefore, Def.(2.7) is the desired analog of the continuum
definition of boundary.

{}From the above examples, it is apparent that there are many simplicial
complexes
which describe the same space. First note that given any simplicial
complex $K$, a {\it subdivision } of $K$ is a simplicial complex $L$
such that $|K|=|L|$ and any simplex in $L$ is contained in a simplex in
$K$. Then two simplicial complexes $K$ and $K'$ are {\it
combinatorially equivalent } if both can be subdivided so that the
simplices of the resulting subdivisions $L$ and $L'$ can be put into
one to one correspondence by relabeling the vertices. For example, the
complexes in Figures 1a) and  1b) are combinatorially equivalent; in
fact Figure 1b) is itself a subdivision of Figure 1a) and thus the
combinatorial equivalence follows trivially. Another example is given
by the simplicial complexes in Figure 3; here both must be subdivided
in order to make the one to one correspondence of the vertices.

The combinatorial equivalence of complexes is related to homeomorphisms
of the simplicial complexes which preserve their linear structure;
since simplicial complexes are polyhedra, PL maps are well defined on
them and such homeomorphisms are PL homeomorphisms.  In fact, one can
easily show that if two simplicial complexes are PL homeomorphic, then
they are combinatorially equivalent. One can also define a more
restricted PL map between simplicial complexes called a  simplicial
map: A {\it simplicial map} $f: K \to K'$ is a PL map that  maps
simplices of $K$  to simplices of $K'$ and is linear on each simplex.
Simplicial maps are more restricted in that they are determined
entirely by their behavior on the vertices of the complex.
Consequently, simplicial homeomorphisms are simply permutations or
relabeling of the vertices of a simplicial complex. Thus one can
restate the definition of combinatorial equivalence of two complexes in
the following way;  $K$ and $K'$ are combinatorially equivalent if
their subdivisions $L'$ and $L'$ are related by a simplicial
homeomorphism.

At this point it is possible begin defining special sets of simplicial
complexes. Historically, a subset of  pure nonbranching simplicial
complexes has been studied because their homology has many properties
similar to that of manifolds. They are given the special name of
pseudomanifolds:\refto{spanier}

\proclaim Definition {(2.8)}. A pseudomanifold $P^n$ is a pure
nonbranching simplicial complex such that i) any two n-simplices can be
connected by a sequence of n-simplices, each intersecting along  some
(n-1)-simplex.

\noindent The reason for the further requirement on pure nonbranching
simplicial complexes is so that  the nth homology group of a
pseudomanifold, $ H_n(P^n)$, has a single generator.  For closed
pseudomanifolds, this condition implies that the ${\bf Z}_2$ homology
always satisfies $ H_n(P^n;{\bf Z}_2)={\bf Z}_2$. This is equivalent to
saying that all closed pseudomanifolds are ${\bf Z}_2$ orientable. In
fact the homology groups yield an equivalent description of closed
pseudomanifolds, namely, a pure nonbranching closed simplicial complex
is a closed pseudomanifold if and only if $ H_n(P^n;{\bf Z}_2)={\bf
Z}_2$.\refto{footnote2} If the simplicial complex is orientable this is
equivalent to $ H_n(P^n)={\bf Z}$. A disadvantage of condition {\sl i)}
is that the boundary of a pseudomanifold is not necessarily itself a
pseudomanifold; the boundary can fail to to be connected and hence fail
to satisfy Def.(2.8). The homology properties of pseudomanifolds that
follow from Def.(2.8) mean that these spaces are very similar to
manifolds; indeed both connected manifolds and conifolds are subsets of
the set of pseudomanifolds. Figures 1a), 3a) and 3b) are all examples
of pseudomanifolds as they are homeomorphic to manifolds.  Figure 4b)
is an example of a pseudomanifold that is not homeomorphic to a
manifold. Finally, note that Figure 2d) is pure and nonbranching but is
not a pseudomanifold as it does not satisfy condition {\sl i)}.

In order to study subsets of pseudomanifolds, three more definitions
are required. These definitions are used to characterize the local
topology of simplicial complexes and thus provide the means for
defining simplicial equivalents of smooth manifolds and conifolds.

\proclaim Definition {(2.9)}. The combinatorial star $St(v)$ of a
vertex $v$ is the complex consisting of all simplices that contain
$v$.

\proclaim Definition {(2.10)}. The combinatorial link $L(v)$ of a
simplex is the subset of the star of $v$ consisting of all simplices in
the star that do not intersect $v$ itself.

\noindent For example, the star of  vertex $f$ in Figure 3a) consists
of the four triangles $ fab$, $ fbc$, $ fcd$, $ fda$ that contain $ f$
and their constituent edges and vertices. The link of this vertex is
the set of vertices and edges $ a$, $b$, $c$, $d$, $ab$, $bc$, $cd$,
$da$ that form a square. Thus the link of $ f$ is homeomorphic to a
circle. Secondly, consider the vertex $ a$ of the pinched torus in
Figure 4b). Its star consists of the six triangles $ acd$, $acb$,
$abc$, $aeg$, $agf$, $ aef$  and their faces.  Its link  consists of
the two disconnected subsets
$ b$, $c$, $d$, $cd$, $bc$, $bd$ and $ e$, $f$, $g$, $ef$, $fg$, $eg$
that are homeomorphic to two disjoint circles.
Thus, the difference in the topology of the neighborhoods of these two
vertices is  carried by the difference in their links. In general it
will be seen that the topology of a neighborhood of a vertex will be
closely related to the properties of its link.

Using Def.(2.1), it is easily verified that the underlying space of the
star $|St(v)|$ is the PL cone over $|L(v)|$ with apex $v$. The
relations between the combinatorial links, stars, and PL cones
 helps to define a simplicial cone of a simplicial complex in the
following way:

\proclaim Definition {(2.11)}. The simplicial cone of a simplicial
complex $K$ with apex $v$, $C(K)$,
is the simplicial complex consisting of the simplicial cones of each
$\sigma \in K$ with apex $v$; that is it consists of all simplices
containing the vertex $v$ and subsets of the vertices of $K$
corresponding to the simplices of $K$.\par

\noindent The simplicial cone of Figure 1a) is a complex consisting of
two tetrahedra and all of their faces as illustrated in Figure 5a).
Similarly, the simplicial suspension of a simplicial complex, $S(K)$,
can be defined by gluing together two simplicial cones $C(K)$ along
their common boundary; an example of a suspension is illustrated in
Figure 5b).
Observe that the PL cone of a k-simplex with apex $v$ is
(k+1)-simplex; for this special case the correspondence of simplicial
cones to PL cones is obvious. In general, the underlying space of the
simplicial cone $C(K)$ is the PL cone over the space $|K|$, that is
$C(|K|)$; in other words $|C(K)|=C(|K|)$.  Similarly, the underlying
space of the simplicial suspension satisfies $|S(K)|=S(|K|)$.  Thus a
simplicial cone is a simplicial representation of the space
corresponding to a  PL cone.

Given these definitions, the simplicial counterparts of various
continuum sets of topological spaces can now be defined. For ease of
presentation, the definitions given below will be for spaces without
boundary. However, note that their generalizations to the case with
boundary are obvious.  First it is very useful to give the simplicial
counterpart of the continuum definition of a homology n-manifold which
is the following:
\proclaim Definition {(2.12)}. A metric space $Q^n$ is a homology
n-manifold if and only if each point $x_0\in Q^n$ has a neighborhood
$N_{x_0}$ which is homeomorphic to a topological cone over a compact
space $L_{x_0}$ which has the same integer homology as $S^{n-1}$.

\noindent Observe that although all manifolds are also homology
manifolds, homology manifolds are not necessarily manifolds; in fact
they are not necessarily conifolds as the compact space $L_{x_0}$ need
not be a closed (n-1)-manifold!  However, there is a close
correspondence between the homology properties of a manifold and those
of a homology manifold. Because of this close correspondence, it is
common to generalize the definition of a homology sphere to include not
only manifolds but homology manifolds with the same integer homology as
a sphere. This convention will be used in Thm.(2.15).

The simplicial counterpart of Def.(2.12) is a subset of
pseudomanifolds:

\proclaim Definition {(2.13)}. A combinatorial homology n-manifold is a
n-pseudomanifold for which the link of every vertex has the same
integer homology as an (n-1)-sphere.

\noindent An example of a combinatorial homology n-manifold is given
later in this section. Although not proven here, one should note that
any simplicial complex which is homeomorphic to a homology manifold as
given by Def.(2.12) satisfies Def.(2.13).

At this point n-manifolds can now be defined:
\proclaim Definition {(2.14)}. A combinatorial n-manifold is a
n-pseudomanifold
for which the link of every vertex is a combinatorial (n-1)-sphere.

\noindent  The main reason for defining combinatorial manifolds in the
above way is that it is a homogeneous definition; namely, no vertex of
the simplicial complex has preferred treatment and the links of the
vertices are all homeomorphic. This is similar to the idea of a
topological manifold where each point has a neighborhood homeomorphic
to a ball. As a necessary and sufficient condition for the star of a
vertex to be a combinatorial n-ball is for the link to be a
combinatorial \hbox{(n-1)}-sphere,
 the definition of a combinatorial n-manifold can be phrased in terms
of the links.
 It is easy to find examples of combinatorial manifolds. For example
figure 3a) is a combinatorial manifold.  The link of every vertex is
topologically a circle. However, Figure 4b) is not a combinatorial
manifold as the link of vertex $ a$ is topologically two disjoint
circles.

It follows immediately from Def.(2.14) that combinatorial manifolds are
homeomorphic to topological manifolds. However, it is important to note
that there are simplicial complexes that are not combinatorial
manifolds that are homeomorphic to topological manifolds as well.  One
can readily construct such a simplicial complex which is a topological
manifold but is not a combinatorial manifold. Consider space $SO(3)/I$
in which all points of the group $SO(3)$ which differ by an element of
the icosahedral group $I$ are identified.  (Recall that the icosahedral
group is the group of symmetries of a icosahedron.) The resulting space
$|\Sigma |$ is a closed smooth 3-manifold because $SO(3)$ is a Lie
group and $I$, being a finite subgroup of $SO(3)$, must act freely.
Furthermore, $|\Sigma |$ is homeomorphic to a combinatorial manifold
$\Sigma $ and has the same integer homology as a 3-sphere. The PL
suspension of $|\Sigma|$, $S(|\Sigma |)$ is not homeomorphic to a
manifold because $|\Sigma |$ is not simply connected; it follows from
the construction of $|\Sigma|$ that the fundamental group is the binary
icosahedral group. This implies that the simplicial suspension
$S(\Sigma) $ is not a combinatorial manifold; in fact it is not
topologically a manifold, but rather is a homology manifold.  However,
the double suspension $S^2(|\Sigma |)$, that is the PL suspension of
$S(\Sigma)$, is homeomorphic to a 5-sphere. This follows from the
double suspension theorem\refto{double} or by explicit construction.
However, even though the space $S^2(|\Sigma |)$ is homeomorphic to a
5-sphere,  the simplicial complex $S^2(\Sigma )$ is not a combinatorial
manifold. In order to see this, note that the apex of either of the two
cones in the second suspension has a link $S(\Sigma )$ which is not a
4-sphere or even a 4-manifold. Thus $S^2(\Sigma)$ does not satisfy
Def.(2.14). Therefore the definition of combinatorial manifold carries
more structure than simply the topology.

The above example raises the issue of how general will a simplicial
complex be if it is homeomorphic to a n-manifold. As just demonstrated,
the links of vertices of such a simplicial complex can be nonmanifolds.
However, it turns out that the links cannot be arbitrary; one can
verify  that the link of every vertex of a simplicial complex
homeomorphic to a n-manifold has the same integer homology as a sphere
even though the link is not necessarily a manifold. The following
theorem gives some necessary conditions on the simplicial complex.

\proclaim Theorem {(2.15)}. Given a connected simplicial complex $K^n$
such that $|K^n|$ is homeomorphic to a closed n-manifold, then $K^n$ is
a combinatorial homology n-manifold. Furthermore, if $n\geq 3$, then
the link of each vertex is also simply connected.

\noindent First, the simplicial complex $K^n$ must be pure; otherwise
there would be points in $K^n$ which have neighborhoods with dimension
less than $n$.  Second, it must be  nonbranching because neighborhoods
of points on (n-1)-simplices at which  three or more n-simplices meet
are not homeomorphic to the interior of a $n$-ball. Hence, $K^n$ is
pure and nonbranching. Since $|K^n|$ is a manifold and all closed
manifolds are ${\bf Z}_2$ orientable, it follows that $H_n(|K^n|;{\bf
Z}_2)={\bf Z}_2$. Therefore, $K^n$ is a closed pseudomanifold.

Next, let $v_0$ be any vertex in $K^n$ and let $U = |K^n| - |St(v_0)|$;
then
 the following relative homology groups satisfy
$$H_*(|K^n|,|K^n| -
\{v_0\})=H_*(|K^n|-U,(|K^n|-\{v_0\}) - U)\eqno(ex1)$$
by the excision
property.\refto{excision} Observe that $|K^n|-U=|St(v_0)|$ and that
$(|K^n|-\{v_0\}) - U= |St(v_0)| -\{v_0\}$ which is homotopic to
$|L(v_0)|$.  Thus it follows from \(ex1) that
$$H_*(|St(v_0)|,|L(v_0)|)=H_*(|K^n|,|K^n|-\{v_0\}).\eqno(ex2)$$
However, $|K^n|$ is a manifold so
$$H_*(|K^n|,|K^n|-v_0)=H_*(B^n,S^{n-1}).\eqno(ex3)$$
Since $B^n$ is
contractible, the exact sequence for relative homology groups implies
that $H_k(B^n,S^{n-1})=H_{k-1}(S^{n-1})$. Hence,
$H_k(|St(v_0)|,|L(v_0)|)=H_{k-1}(S^{n-1})$. Using the exact sequence
again with the contractibility of $|St(v_0)|$, it follows that
$H_k(|St(v_0)|,|L(v_0)|)=H_{k-1}(|L(v_0)|)$. Hence,
$H_*(|L(v_0)|)=H_*(S^{n-1})$. Therefore, $K^n$ is a combinatorial
homology manifold.

Finally, let $v_0$ be any vertex in $K^n$ where $n\geq 3$, then
$C(|L(v_0)|)$ is a manifold by assumption. Observe that if $M^n$ is any
manifold with $n\geq 3$, then $\pi_1(M^n-\{p \})=\pi_1(M^n)$ for point
$p\in M^n$. This is due to the fact that any curve can be moved around
an isolated point in three or more dimensions without intersecting the
point. Hence, $\pi_1(C(|L(v_0)|)-\{a\})=\pi_1(C(|L(v_0)|))=1$ where $a$
is the apex of the cone.  Furthermore,
$\pi_1(C(|L(v_0)|)-\{a\})=\pi_1(I\times |L(v_0)|)=\pi_1(|L(v_0)|)$.
Therefore, $\pi_1(|L(v_0)|)=1$. Q.E.D.

Finally, the simplicial counterparts of topological conifolds  as
defined in Def.(1.1) can be presented. By analogy with  Def.(2.14) for
a combinatorial manifold, combinatorial conifolds are defined as
follows:

\proclaim Definition {(2.16)}. A combinatorial n-conifold is a
n-pseudomanifold for which the link of every vertex is a closed
connected combinatorial (n-1)-manifold.

\noindent  Clearly, all of the singular points of a combinatorial
conifold are a subset of the set of vertices of the simplicial complex.
Hence, a combinatorial conifold is a manifold everywhere except
possibly at a countable set of vertices. This  parallels the definition
of a topological n-conifold,
 for which the neighborhoods of all but a countable set of points are
homeomorphic to n-balls. The requirement that the links of vertices be
combinatorial (n-1)-manifolds is the natural extension of the
requirement that the links be combinatorial (n-1)-spheres in the
definition of combinatorial manifold.  The class of combinatorial
n-conifolds quite clearly includes all combinatorial n-manifolds by
definition; again as in the continuum case, the class of combinatorial
n-conifolds differs from that of n-manifolds only for $n\ge 3$. Thus
Def.(2.16) is the logical analog of Def.(2.14).  Figure 6b) provides an
example of a combinatorial conifold that is not a combinatorial
manifold.

As for combinatorial manifolds, a connected combinatorial  conifold is
a pseudomanifold by definition. However, again the converse is not
true.  The simplicial suspension of any n-pseudomanifold will be a
(n+1)-pseudomanifold; for example the simplicial suspension of the
pinched torus in Figure 5b) is a 3-pseudomanifold.  However, in general
such suspensions will not be combinatorial conifolds as the links of
vertices need not be connected manifolds. Indeed, the link of the apex
of the suspension of Figure 4b)  is clearly not a manifold.
Furthermore, all simplicial complexes that are topologically
n-conifolds are not necessarily combinatorial n-conifolds. This follows
immediately from the fact that the set of combinatorial n-conifolds
includes all combinatorial n-manifolds. Again the double suspension
$S^2(\Sigma)$ provides an easily understood example; recall that the
links are not combinatorial 4-manifolds. Therefore, $S^2(\Sigma)$ fails
to be a combinatorial 5-conifold as well as a combinatorial 5-manifold.
It is easy to persuade oneself that a similar construction can be done
to form more general examples of simplicial complexes that are
topologically  conifolds but fail to be combinatorial conifolds.  Thus,
combinatorial conifolds, like combinatorial manifolds, are simplicial
counterparts of topological conifolds with additional nice structure.
The precise nature of this nice structure will be discussed in the next
section.

\head{3.~Triangulation of Manifolds and Conifolds}

Obviously, combinatorial manifolds and conifolds as defined in
Def.(2.14) and Def.(2.16)
are all topological manifolds and conifolds. However, as seen in the
last section, not all simplicial complexes that are topological
manifolds and conifolds are actually combinatorial manifolds and
conifolds. Finally, a priori, it is not clear what the relationship is
between these combinatorial spaces and the corresponding smooth
versions.  Therefore, an explicit characterization of the connection of
smooth manifolds and conifolds to their combinatorial counterparts is
desirable and necessary.  Indeed one anticipates that a close
connection between smooth and combinatorial spaces exists; very much
like smooth spaces, combinatorial spaces have a nice structure that
will enable integration and differentiation to be well defined as
necessary for physical applications. It turns out  that all smooth
manifolds and conifolds have combinatorial counterparts and in less
than seven dimensions, any combinatorial manifold or conifold
corresponds to a unique smooth manifold or conifold respectively.

In order to discuss the connection of smooth spaces to combinatorial
spaces, the first concept needed is that of a triangulation. Given any
topological space $P$, a {\it triangulation} consists of a simplicial
complex $K$ and a homeomorphism $t:|K|\rightarrow P$. One can show that
all polyhedra as defined in Def.(2.2) admit a triangulation. Moreover,
any topological space that admits a triangulation is homeomorphic to a
polyhedron. Therefore spaces that admit a triangulation are nice in the
sense that they have the same properties as polyhedra.  A {\it
combinatorial triangulation} of a manifold $M^n$ consists of a
combinatorial manifold $K^n$ and a homeomorphism $t:|K^n|\rightarrow
M^n$. Similarly, a combinatorial triangulation of a conifold $X^n$
consists of a combinatorial conifold $K^n$ and a homeomorphism
$t:|K^n|\rightarrow X^n$.

It is important to note that not all triangulations of manifolds are
combinatorial triangulations as was illustrated with the $S^2(\Sigma)$
example of the previous section. Such non-combinatorial triangulations
of manifolds are referred to as {\it weak triangulations}.
However, note that a direct consequence of Thm.(2.15) and the observed
properties of manifolds and pseudomanifolds is that all weak
triangulations of closed n-manifolds for $n\le 3$ are in fact
combinatorial triangulations. In one dimension, this result follows by
construction. In two and three dimensions, recall that by Thm.(2.15),
the links of a weak triangulation of a n-manifold must be homology
spheres.  However, in dimensions one and two, a homology sphere is a
combinatorial sphere; this observation is trivial in one dimension and
in two dimensions follows immediately by recognizing that the only
2-pseudomanifold with the same integer homology as a 2-sphere is the
2-sphere itself.\refto{later} Therefore, in one, two and three
dimensions, all weak triangulations are in fact combinatorial
triangulations. The example of the 5-sphere given in section 2 already
shows this result fails in five dimensions and it is unsolved for
4-manifolds.

The first task is to show that every smooth manifold and conifold admit
combinatorial triangulations.\refto{cairns} Intuitively, one should
expect this to be true because these spaces admit a smooth atlas by
definition and it seems that by judicious choice, the smooth charts can
be taken to correspond to the simplices of a combinatorial manifold.

In order to prove that all smooth n-manifolds have triangulations,
assume that $M^n$ is a closed smooth n-manifold. By a standard
embedding theorem it can be smoothly embedded in ${\bf R}^{2n+1}$ as a
closed subset. Note that $M^n$ inherits a metric and curvature from the
embedding. Next observe that ${\bf R}^{2n+1}$ has a nice family of
triangulations consisting of  combinatorial manifolds $K$ and a
simplicial homeomorphism; i.e. this family of triangulations consists
of tessellating ${\bf R}^{2n+1}$ with simplices.  Pick one of these
triangulations  of ${\bf R}^{2n+1}$ such that the size of each simplex
in the image of $K_0$, $Im(K_0)$, is small compared to the curvature of
$M^n$; in other words, choose the simplices of the triangulation to be
small enough that the manifold appears to be approximately flat inside
the simplices. Observe that $M^n$ intersects $Im(K_0)$ and the
triangulation can be chosen such that $M^n$ is in general position with
respect to the n-simplices of $Im(K_0)$. This is easy to see as if
$M^n$ does not intersect $Im(K_0)$ in general position for the initial
choice of triangulation, the vertices of $Im(K_0)$ can be moved so that
it does. But moving the vertices corresponds to a simplicial
homeomorphism, i.e. to another nice triangulation, therefore one can
choose this triangulation for ${\bf R}^{2n+1}$ from the start. Since it
is in general position with respect to $M^n$, no interiors of simplices
in $Im(K_0)$ with dimension less than (n+1) intersect $M^n$. This
result combined with the fact that the size of the simplices of
$Im(K_0)$ is small compared to the curvature of $M^n$ means that each
(n+1)-simplex which has non-empty intersection with $M^n$ intersects it
in a unique point which is in the interior of the (n+1)-simplex.
Furthermore, if the simplices are chosen small enough, then the
intersection of $M^n$ with each (2n+1)-simplex is a convex n-ball. The
collection of all of these n-balls will yield curved simplices which
correspond to the image of a simplicial complex
$K^n$ homeomorphic to $M^n$.  Thus any closed smooth n-manifold has a
combinatorial triangulation.  Observe that the homeomorphism is
actually a smooth map on the interior of every simplex by construction;
therefore, this triangulation of the n-manifold is particularly well
behaved.

In order to prove that smooth manifolds that are not compact have
triangulations, note that the compactness of $M^n$ is used in picking
the triangulation of ${\bf R}^{2n+1}$ to be small relative to the
curvature of the manifold. If the manifold is not compact, the
curvature may become larger as one moves toward infinity. However, note
that the triangulation can be chosen such that the images of the
simplices are shrinking as a function of distance. Then the rest of the
above argument for closed n-manifolds follows through. A similar
technique also applies in the case of a smooth manifold with boundary.
Thus, in all cases the above method produces a combinatorial
triangulation of the smooth manifold.

Given the result that all smooth n-manifolds have combinatorial
triangulations, it immediately follows that all smooth n-conifolds
$X^n$ have them too.  Recall that the singular set $S$ of a n-conifold
is the set of points whose
neighborhoods are not homeomorphic to the interior of a cone over a
(n-1)-sphere.\refto{I} Delete conical neighborhoods of the singular
set $S$ and then triangulate the resulting smooth manifold with
boundary, $X^n-N(S)$. Finally, take a simplicial cone as given in
Def.(2.11) of each boundary (n-1)-manifold to yield a combinatorial
triangulation of the smooth n-conifold $X^n$.  Therefore,  any smooth
manifold or conifold has a combinatorial triangulation; that is every
smooth manifold or conifold has a combinatorial counterpart.

The next step is to find the conditions for which the converse of the
above statement is true; that is under what conditions does a
combinatorial manifold or conifold have a smooth counterpart. In order
to do so, it is useful to introduce a more general set of manifolds
than smooth manifolds, namely PL manifolds.\refto{rs}
\proclaim Definition {(3.1)}.  A topological manifold $M^n$ is a  PL
manifold if and only if there is an atlas
$\{(U_\alpha,\varphi_\alpha)\}_{\alpha\in \Lambda}$ such that the
mapping
$$\varphi_\beta
\varphi_\alpha^{-1}: \varphi_\alpha(U_\alpha \cap U_\beta) \to
\varphi_\beta(U_\alpha \cap U_\beta)$$
is a PL mapping between subsets of ${\bf R}^n_+$.\par
\noindent  Note
that there is no ambient embedding space needed in the definition of
a PL manifold, in contrast to the definition of a combinatorial manifold.
Additionally,
\proclaim Theorem {(3.2)}. Combinatorial manifolds are equivalent to PL
manifolds.\par
\noindent Let $K^n$ be a combinatorial n-manifold; then $|K^n|$ is a PL
manifold with $K^n$ as its triangulation. This can be demonstrated by
defining a collection of neighborhoods $U_v = St(v) - L(v)$ for each
vertex $v$ in $K^n$. These charts cover $|K^n|$ and since each $U_v$ is
an open subset of the polyhedron $|K^n|$, it follows that each $U_v$ is
itself a polyhedron. Furthermore, each $U_v$ is PL homeomorphic to the
interior of a standard n-simplex $\sigma^n$ as $K^n$ is a combinatorial
n-manifold. Choose a set of PL homeomorphisms $\varphi_v:U_v\to
|\sigma^n|$ to be such maps. Next the intersections $U_v\cap U_w$ are
also polyhedra which are PL homeomorphic to an open set contained in
the interior of $|\sigma^n|$. Furthermore, the overlap maps of the
intersection are PL homeomorphisms as given in Def.(3.1). Therefore
$|K^n|$ is a PL manifold with triangulation $K^n$ by construction.

Conversely, given any PL manifold $M^n$, it has a combinatorial
triangulation $K^n$. The following lemma, proven in Appendix A, is
needed in order to prove this:
\proclaim Lemma {(3.3)}. Let $P_1$ and $P_2$ be polyhedra,
$S_1 \subseteq P_1 $ and $S_2 \subseteq P_2$ be subpolyhedra and
$\psi: S_1 \to S_2$ be a PL homeomorphism. Then $P = P_1 \cup_\psi P_2$
is a polyhedron where $ P_1 \cup_\psi P_2$ is the disjoint union of the
two polyhedra with points $x_1\in S_1$ and
 $x_2 \in S_2$ identified if and only if $\psi (x_1) = x_2$.\par
\noindent Denote the atlas of $M^n$ by $\{U_\alpha,
\varphi_\alpha\}_{\alpha\in \Lambda}$. Each set
$\varphi_\alpha(U_\alpha) \subseteq {\bf R}^n$ is a polyhedron because
it is an open subset of ${\bf R}^n$. Next $\varphi_\alpha(U_\alpha \cap
U_\beta) \subseteq \varphi_\alpha(U_\alpha)$ and
$\varphi_\beta(U_\alpha \cap U_\beta) \subseteq \varphi_\beta(U_\beta)$
are open subsets and therefore subpolyhedra. Moreover, each map
$\psi_{\alpha\beta}:\varphi_\alpha(U_\alpha \cap U_\beta)\to
\varphi_\beta(U_\alpha \cap U_\beta)$ where $\psi_{\alpha\beta}=
\varphi_\beta\varphi_\alpha^{-1}$ is a PL homeomorphism. The above
lemma implies that $\varphi_\alpha(U_\alpha )\cup_{\psi_{\alpha\beta}}
\varphi_\beta( U_\beta)$ is a polyhedron. Thus for any $U_\alpha$ and
$U_\beta$, the above construction produces another polyhedron that is
the disjoint union of the two spaces. Next note that this construction
can be repeated between the neighborhood $U_\beta$ and another
neighborhood $U_\gamma$ to form a new polyhedron
$\varphi_\beta(U_\beta)\cup_{\psi_{\beta\gamma}}
\varphi_\gamma(U_\gamma)$. Furthermore, this polyhedron can be joined
to the first to form another polyhedron by noting that the polyhedron
$\varphi_\beta(U_\beta)$ is common to both. Finally, by successively
adding the remaining neighborhoods $U_\alpha$ in this fashion, it
follows that the resulting space $P$ is a polyhedron homeomorphic to
$M^n$. Furthermore all polyhedra have a triangulation so $P= |K^n|$
where $K^n$ is a simplicial complex and  hence $M^n = |K^n|$.  Finally
the star of every vertex of $K^n$ is PL homeomorphic to a n-ball
because each point in $M^n$ has a neighborhood PL homeomorphic to a
subset of ${\bf R}^n$. Therefore, $K^n$ is a combinatorial n-manifold
and thus $M^n$ has a combinatorial triangulation. Moreover, this
constructive proof shows that the map $t:|K^n|\to M^n$ is PL in the
sense that $\varphi_\alpha t$ is a PL homeomorphism for all charts in
the atlas. Q.E.D.

Thus  PL manifolds can be thought of as coordinate independent
representations of combinatorial manifolds. Now,
the result that every smooth n-manifold has a combinatorial
triangulation immediately implies that all smooth n-manifolds have a PL
counterpart in any dimension. Moreover, the correspondence of
combinatorial manifolds to smooth manifolds can be addressed using the
known results from topology for the correspondence of PL manifolds to
smooth manifolds.\refto{toy}

In order to understand the results on this correspondence, recall that
the structure group of the tangent bundle of a smooth manifold is
$GL(n,{\bf R})$. This follows from the fact that the  maps on the
overlaps of the charts of the smooth manifold are diffeomorphisms on
${\bf R}^n$ and diffeomorphisms act on vectors in ${\bf R}^n$ by
general linear transformations.
In fact, without loss of generality, by appropriately choosing a
Riemannian metric so that orthonormal vectors are defined, one can
assume that the structure group is $O(n)$. The structure group of the
tangent bundle over a PL manifold is not so simple; it is the group of
PL homeomorphisms, $PL(n)$.  Observe that the structure groups on the
tangent bundles carry the information about the structure (PL or
smooth)  of the base manifold. Therefore the question of whether or not
a given PL manifold admits a smoothing is equivalent to the question of
whether or not one can change the structure group on the given PL
tangent bundle from $PL(n)$ to $O(n)$.

Whether or not there is an obstruction to placing an $O(n)$ structure
group on a given PL tangent bundle is determined by the cohomology of
the PL manifold:\refto{toy}

\proclaim Theorem {(3.4)}. Let $M^n$ be a PL manifold and $\Gamma_k$ be
the group of diffeomorphisms of the k-sphere, $f:S^k \to S^k$, modulo
those which extend to diffeomorphisms of the (k+1)-ball, $f: B^{k+1}
\to B^{k+1}$. Then $M^n$ is smoothable if and only if  the obstructions
$c_k(M^n) \in H^{k+1}(M^n; \Gamma_k)$ vanish for  $0\le k\le n-1$.\par
\noindent Note that $\Gamma_k$ is the same as the set of inequivalent
smooth structures on  a PL k-sphere. If the obstructions vanish, than
there is at least one smoothing; that is
 one can find a smooth atlas on $M^n$ that is diffeomorphic its  PL
 atlas when restricted to the interior of each neighborhood.  If they
 do not vanish, than there is no smoothing of the given PL manifold.
The number of smoothings of a PL manifold that is smoothable is
determined by the following\refto{toy}

\proclaim Theorem {(3.5)}. A PL homeomorphism $f:M^n \to N^n$ is
equivalent to a smooth map if the obstructions of Thm.(3.4) vanish and
$c_k(f) \in H^{k+1}(M^n; \Gamma_{k+1})$ vanish for  $0\le k\le n$.\par
\noindent These cohomology conditions are necessary and sufficient in
any dimension although note that the results on smoothing 2-manifolds
and 3-manifolds can be proven independently.

Using these results, it can be proven that every PL manifold in less
than seven dimensions has a unique smoothing; this follows from the
fact that $\Gamma_k = 0$ for $k\le 6$.  In seven dimensions, all PL
manifolds have smoothings, but one can show that there are PL manifolds
that do not have a unique smoothing. In eight or more dimensions, one
can show that there are both
PL manifolds that do not correspond to smooth manifolds and PL
manifolds that do not have a unique smoothing. Therefore,  PL manifolds
are a more general set of topological manifolds than smooth manifolds;
however in less than seven dimensions, PL manifolds have a unique
correspondence to smooth manifolds.  Thus in less than seven
dimensions, combinatorial manifolds uniquely correspond to smooth
manifolds.  This connection between smooth manifolds and PL manifolds
is the reason that combinatorial triangulations are preferred over weak
triangulations for the purposes of this paper.\refto{2ndreason}

Similarly, the conditions under which a combinatorial conifold has a
smooth counterpart are best discussed in terms of PL conifolds.  PL
conifolds are defined by requiring that the topological n-conifold  of
Def.(1.1) admit a PL atlas;

\proclaim Definition {(3.6)}. A PL atlas on a n-conifold $X^n$ is a
collection $\{(U_\alpha,\varphi_\alpha)\}_{\alpha\epsilon\Lambda}$ of
open sets and homeomorphisms indexed by a set $\Lambda$ satisfying the
following:
\item{i)} The sets $U_\alpha$ cover $X^n$.
\item{ii)} $X^n-S=\hbox{\lower.9ex\hbox{${\bigcup
\ \ }\atop{\alpha\in \Lambda_0}$}}U_\alpha$ for some subset
${\Lambda}_0\subset{\Lambda}$.
\item{iii)} For $\alpha\in {\Lambda}_0$, $\varphi_\alpha$ is a PL
homeomorphism of $U_\alpha$ to an open set in ${\bf R}^n_+$.
\item{iv)} For each $\alpha\in \Lambda -{\Lambda}_0$, $U_\alpha$ is a
conical neighborhood of a singular point and $\varphi_\alpha$ is a PL
homeomorphism onto the interior of a cone.\par

\noindent Again, it can be proven that
 PL conifolds are equivalent to combinatorial conifolds.  This result
can be seen easily from the corresponding result for manifolds,
Thm.(3.2): Let $K^n$ be a combinatorial n-conifold; then it follows
immediately by the same arguments as given in Thm.(3.2) that $|K^n|$ is
a PL conifold with $K^n$ as its triangulation. Next let $X^n$ be a PL
conifold.  Excise neighborhoods of all the singular points, $N(S)$, of
$X^n$; the result is a PL manifold with boundary. Next observe that by
Thm.(3.2), the PL manifold is equivalent to a combinatorial manifold.
Additionally the neighborhood of each singular point is a PL cone over
a (n-1)-manifold and is easily seen to be PL homeomorphic to a
simplicial cone over the manifold. Thus $X^n = |K^n|$ where $K^n$ is a
simplicial complex. Finally, as each piece in the construction of the
simplicial n-conifold $K^n$ is combinatorial, the conifold itself is
combinatorial.  Thus PL conifolds can be thought of as coordinate
independent representations of combinatorial conifolds.

As in the manifold case, PL conifolds are closely related to smooth
conifolds; again it follows immediately that every smooth n-conifold
has a PL counterpart in any dimension.  Furthermore, the results for
smoothing PL manifolds can be extended to prove similar results for PL
conifolds although the details will not be presented here.  In
particular, it can be proven that
any PL conifold of dimension less than seven has a unique smoothing.
Thus in  fewer than seven dimensions, smooth manifolds and conifolds
have a unique correspondence to combinatorial or equivalently PL
manifolds and conifolds. Moreover, it is precisely this connection that
allows for a concrete discussion of the issues involved in summing over
physically distinct manifolds and conifolds to be formulated in terms
of combinatorial manifolds and conifolds.

The final issue is the relation of topological n-manifolds and
n-conifolds to smooth n-manifolds and n-conifolds. Although only smooth
spaces are relevant to physics, it is useful to understand their
relationship with topological spaces as it characterizes the additional
structure that smooth spaces carry.  Two smooth structures are said to
be equivalent if they are diffeomorphic to each other. Similarly, two
PL structures are equivalent if they are PL homeomorphic to each other.
Clearly, by the previous discussion, the number of inequivalent PL
structures  on a given topological manifold or conifold determines the
number of inequivalent smooth structures and there is a unique
correspondence between the numbers in less then seven dimensions.  Thus
results on PL structures  and smooth structures are interchangable in
less than seven dimensions.

It turns out that whether or not all smooth structures on a manifold
are equivalent depends on dimension.  In dimension three or less, every
topological manifold admits a smooth structure.\refto{smoothstructure}
Furthermore, one can prove that this smooth structure is unique.
Therefore, in dimension three or less, there is no difference between
topological, PL and smooth manifolds.  In more than five dimensions,
the number of inequivalent PL structures can be characterized in a
manner similar to that used in the smoothing of PL manifolds. Whether
or not a given topological manifold $M^n$ admits a PL structure is
determined by whether or not the structure group of the topological
manifold, $Top(n)$, can be replaced by $PL(n)$. The obstruction to
placing a PL structure on a given topological manifold $M^n$ is
determined by the following:\refto{toy}

\proclaim Theorem {(3.7)}. Let $M^n$ be a topological manifold with
$n\ge 5$. Then $M^n$ has a smooth structure if and only if the
invariant $ks(M^n) \in H^4(M^n, \partial M^n; {\bf Z}_2)$ satisfies
$ks(M^n) = 0$.  Furthermore, given  a continuous homeomorphism $h:M^n
\to N^n$ between PL manifolds, it is equivalent to a PL homeomorphism
if and only if the invariant $ks(h) \in H^3(M^n, \partial M^n; {\bf
Z}_2)$ satisfies $ks(h)=0$. \par
Using these results, it can be proven that there are topological
n-manifolds that do not admit a PL structure in five or more
dimensions.  In addition, it follows from the above theorem
that there are a finite number of inequivalent PL structures on all
n-manifolds in five or more dimensions if the cohomology is finitely
generated. In particular, all compact n-manifolds in five or more
dimensions have finitely generated cohomology and thus admit a finite
number of smooth structures.

Thm.(3.7) breaks down  in four dimensions. It turns out that the
vanishing of $ks(M^n)$  is only a necessary condition in four
dimensions; it is not sufficient. This is the reason that many issues
concerning smooth structures on 4-manifolds remain open. However, there
are several important results on 4-manifolds that have recently been
proved.
First, as shown by Freedman,\refto{freedman} there are topological
4-manifolds that admit no PL structure. Even worse, one can show that
there are topological 4-manifolds  that do not even admit a weak
triangulation from results of Donaldson.\refto{donaldson} Therefore,
there are topological 4-manifolds that cannot be realized in terms of a
simplicial complex.  A discussion of such a topological 4-manifold,
$||E8||$, is provided in Appendix B.  In addition, it is well known
that some 4-manifolds admit more than one smooth structure; in fact,
there are an infinite number of inequivalent smooth structures on
certain 4-manifolds. For compact 4-manifolds, it can be shown that the
number is countably infinite; for example, the 4-manifold $CP^2\#
9(-CP^2)$, a connected sum of complex projective space with nine copies
of itself with the opposite orientation, has a countably infinite
number.\refto{toy} Even more interesting is the result that the number
of different smooth structures is uncountable for certain open
manifolds. In particular, ${\bf R}^4$ and ${\bf R}\times S^3$ both have
an uncountable number of distinct smooth structures.\refto{freedman}
This is shockingly different than the case for ${\bf R}^n$ in any other
dimension; by Thm.(3.7), all other ${\bf R}^n$ have a unique PL
structure as their cohomology vanishes. Moreover they have a unique
smooth structure by Thm.(3.4) and Thm.(3.5) for the same reason. It is
clear that the issue of smooth structures on manifolds in four
dimensions is much more complicated than in any other dimension.

Similar results on smooth structures apply to n-conifolds; these
results are summarized below as the details are not directly relevant
to this paper.\refto{bigmath} All conifolds are manifolds in dimensions
one and two and therefore have a unique smooth structure in these
dimensions. The result that 3-manifolds have a unique smooth structure
implies that  3-conifolds do as well.  By removing conical
neighborhoods around singular points of the conifold, one obtains a
manifold with boundary. This manifold has a unique smooth structure; by
gluing back the conical neighborhood of the singular points, one
produces a unique smooth structure on the conifold. Consequently, there
is no difference between topological, smooth and combinatorial
conifolds in dimension three or less.  In four or more dimensions,
n-conifolds may admit more than one inequivalent smooth structure or
may admit no smooth structure.  In addition to the obvious examples of
n-conifolds that are topological manifolds that do not admit a PL
structure, one can show that the suspension of the 4-manifold $||E8||$
will be a topological conifold which does not have a PL structure.
Therefore, there are topological conifolds besides those that are also
topological manifolds that do not admit PL structures.

The unique correspondence of smooth manifolds and conifolds to
combinatorial manifolds and conifolds in less than seven dimensions
implies that the information about the smooth structure of the space is
carried by its combinatorial triangulation. Indeed, equivalence of
smooth structures can also be reexpressed in terms of the properties of
the combinatorial spaces; combinatorially equivalent n-manifolds have
both the same topology and equivalent smooth structures. Similarly, two
combinatorial n-conifolds that are combinatorially equivalent have both
the same topology and equivalent smooth structures.  Thus it follows
from the results on smooth structures that all combinatorial manifolds
that are topologically equivalent are actually combinatorially
equivalent in three or fewer dimensions.  In four dimensions, it
follows that there are an infinite number of combinatorially
inequivalent triangulations of some closed 4-manifolds such as $CP^2\#
9(-CP^2)$ as combinatorially inequivalent triangulations correspond to
distinct smooth structures.  Similarly, the result that there are an
uncountable number of inequivalent smooth structures on ${\bf R}^4$
implies that there are an uncountable number of combinatorially
inequivalent triangulations of it.  In dimensions five and six,  there
are a finite number of combinatorially inequivalent triangulations of
some n-manifolds as there are a finite number of distinct smooth
structures.  In dimension seven or greater, as combinatorial manifolds
no longer necessarily correspond to smooth manifolds, a combinatorial
manifold no longer necessarily specifies a unique smooth structure on
the manifold. Finally, similar results for combinatorial n-conifolds in
less than seven dimensions follow from the corresponding results on
smooth n-conifolds.

These first two sections provide the tools to study the topological
issues involved formulating quantum amplitudes involving sums over
smooth spaces such as \(fpartfn).  Since the simplicial complexes are
discrete descriptions of smooth manifolds and conifolds, the question
of the algorithmic decidability of manifolds and conifolds can be
phrased using them. Furthermore, the unique correspondence of smooth
manifolds and conifolds and their combinatorial counterparts in less
than seven dimensions  provides the means of taking into account
inequivalent smooth structures in a sum over topological spaces.

\head{4.~Classifiability and Decidability of Manifolds and Conifolds}

The canonical rule in the formulation of quantum amplitudes via a sum
over histories is that only physically distinct histories are included
in the sum. As discussed in section 2 of  part I (Ref.[\cite{I}]), a
history in Euclidean gravity consists of both a topological space such
as a manifold or a conifold and a metric; thus physically distinct
histories must consist not only of physically distinct metrics but also
physically distinct topological spaces.  Therefore, one needs to have a
method for determining whether or not two topological spaces are
physically distinct. Having an abstract definition of the set of
topological spaces is not actually enough; the existence of such an
abstract definition does not ensure that the set can be constructed.
The issue of whether or not there is a finite procedure for determining
if a given topological space actually satisfies the abstract definition
of the set
is called algorithmic decidability. Secondly, an abstract definition
of the set does not ensure that one can find a unique collection of
representatives. The issue of whether or not there is a finite
procedure for determining if a given member of the set is distinct from
another member of the set is called classifiability. Therefore the
vital question at hand is whether manifolds and/or conifolds are
algorithmically decidable and classifiable.

As mentioned in part I, any finite set  has an algorithmic
description.  Thus the issue of whether or not a set is algorithmically
decidable is only of interest in the case of infinite sets.  The sets
of compact connected combinatorial n-manifolds and compact
combinatorial n-conifolds are countably infinite:  A compact connected
manifold contains a finite number of n-simplices. Since one can obtain
at most a countable number of different spaces by gluing together a
finite number of n-simplices, it follows there are at most a countable
number of compact connected combinatorial n-manifolds. (One should
observe that the compactness of the manifolds is why there are only a
countable number of manifolds. In fact, in three or more dimensions
there are an uncountable number of n-manifolds which are not compact.)
Similarly  the set of compact connected combinatorial n-conifolds is
countably infinite. Therefore,  finite algorithms can be expressed in
terms of this finite set of elements in a given dimension n. Thus there
is the possibility that there may exist  finite algorithms for the
algorithmic decidability and classifiability of compact connected
 n-manifolds and n-conifolds. It turns out that the existence of such
algorithms depends on dimension.

\subhead{\undertext{4.1~Algorithmic Decidability and
Classifiability}}

It is useful to discuss certain examples of algorithmic decidability
and classifiability before discussing the particular case of
n-manifolds and n-conifolds.  First, note that there are examples of
infinite sets where no algorithms exist.  One can make the simple
observation that if a set is uncountable, then there is no way that one
can describe it algorithmically because there is no way to place the
members of the set in one to one correspondence with the integers to
facilitate the development of any finite procedure. Thus simple
examples of sets that do not have algorithmic descriptions are the set
of all subsets of the integers and the set of real numbers. Therefore
the questions of algorithmic decidability and classifiability are
nontrivial only in the case of countably infinite sets. It turns out
there are examples with and without finite algorithms for this case.

An example of an infinite set which can be described algorithmically is
the set of prime numbers. Given any natural number $N$ one can write an
algorithm which will decide whether or not it is prime, namely, test
whether or not any integer between $1$ and $N$ divides $N$: Start with
the number $2$. If $2$ divides $N$, then $N$ is not prime and the test
stops.  However, if $2$ does not divide $N$, then the same test is
repeated
for each consecutive integer until either the test stops or the
integer equals  $N$. If $N$ is reached before the test stops then $N$
is prime. This is quite clearly a finite algorithm for deciding whether
or not an integer is prime.

An important example of an infinite algorithmically decidable set is
the set of all finitely presented groups.\refto{stillwell} Such
finitely presented groups are  very important for topology as they
correspond to the fundamental groups of manifolds.
A finitely presented group is a group with a finite number of
generators and relations.  A generator is represented by  $a_i^{}$ and
has an inverse $a_i^{-1}$.  Elements of the group correspond to finite
products of these generators, e.g. $a_1^{},\  a_2^{}a_1^{-1}a_2^{},\
a_3^2 a_2^{} a_1^{-1}$, and are called {\it words}. The identity
element $1$ of the group is the empty word, that is the word containing
no generators.
A relation $r_i=1$ sets particular words of the group equal to the
identity element. Two words $w_1$ and $w_2$ are said to be equivalent
if one can be transformed into the other by a finite sequence of
insertions or deletions of the relations or of the trivial relations
$a_i^{}a^{-1}_i = a^{-1}_i a_i^{} = 1$.  The presentation of a finitely
presented group is given by \hbox{$<a_1^{},a_2^{},\ldots,a_n^{};
r_1,r_2,\ldots r_n>$} which denotes the set of all equivalence classes
of finite words in the generators $a_i^{}$ with respect to the
relations $r_i$.
For example one presentation of the group $\bf Z$ is $< a;->$
consists  of one generator $a$ and no relations. Similarly, one
presentation of ${\bf Z}_2 = <a;a^2>$ consists of one generator and one
relation and one presentation of the permutation group on three objects
\hbox{ $P= <a_1^{},a_2^{}; a_1^3,\  a_2^2, \
a_1^{}a_2^{}a_1^{-2}a_2^{-1}>$} consists of two generators and three
relations. It is clear from its definition that the set of all finitely
presented groups is algorithmically decidable.

Although it is easy to algorithmically describe what a finitely
presented group is, note that distinct sets of generators and relations
may actually be different presentations of the same group.
For example, the presentations \hbox{ ${\bf Z} = <a_1^{}, a_2^{};
 a_2^{}> $}, \hbox{${\bf Z}_2 = <a;a^2,\ a^4>$} and \hbox{$P=
<a_1^{},a_2^{}; a_1^3,\ a_2^2,\ a_1^{}a_2^{}a_1^{}a_2^{-1}>$} are all
different presentations of the groups $\bf Z$, ${\bf Z}_2$ and $P$.
Note that both the number of generators and number of relations can
differ in different presentations of the same group. Therefore the set
of all finitely presented groups contains more than one representative
of the same group.

Clearly it would be useful to eliminate this redundancy and find the
set of unique representatives of all finitely presented groups.  Such a
set is a classification of the set of all finitely presented groups.
However,
there is no finite algorithm for finding unique representatives of
such groups.  As finite presentations of the same group can have
different numbers of generators and relations as well as different
relations between the generators, there is no way to determine whether
or not a given finite presentation is equivalent to a fiducial
presentation by comparing generators and relations alone. One instead
has to determine whether or not the given finite presentation generates
the same group as the fiducial one by a finite algorithm.
In order to have such an algorithm, one must have a finite algorithm
for determining whether or not an arbitrary word $w_i$ is equivalent to
$1$. This problem is called {\it the word problem for finitely
presented groups} and one can prove that there is no solution to
it.\refto{wordproblem} An important point is that one can show that
there is no solution to the word problem for particular presentations
of groups, not just for the set of all finite presentations of groups.
There are several known examples of such
groups;\refto{boone,post,examples} one such finite  presentation of a
group is given in Appendix C.  The fact that there are explicit finite
presentations that can be proven to be unsolvable emphasizes  the point
that the unsolvability of the word problem is not a property of the set
but rather of a particular
finite presentation of a group. Thus the issues raised by the
unsolvability of the word problem inevitably arise whenever these
particular presentations appear, not just with the set of all finite
presentations of groups in its entirety.

Finally, another related problem which has no algorithmic solution is
{\it the isomorphism problem for finitely presented groups}, i.e. there
is no finite
algorithm to prove whether or not two arbitrary presentations are
isomorphic groups.\refto{rabin} Intuitively, this is related to the
word problem because in order to prove that two presentations generate
isomorphic groups, one must first identify which elements are trivial
in each presentation. In fact, one can prove that the two problems are
completely equivalent.\refto{isoproblem} The unsolvability of the word
problem and the isomorphism problem will be seen to be directly
relevant to the algorithmic decidability and classifiability of
manifolds.

These examples illustrate the basic procedure for determining whether
or not a countably infinite set is algorithmically decidable or
classifiable.  That is, one begins with an algorithmically decidable
set that includes all members of the set of interest; i.e. the set of
all positive integers or the set of all finitely generated groups.
Then one asks whether or not there is an algorithm to select out the
set of elements satisfying the more restrictive definition; i.e. the
set of all primes or the set of all unique representatives of groups.
Finally the important part of the procedure is not the initial
algorithmically decidable set (so long as it contains all members of
the subset of interest) but whether or not there is an algorithm for
selecting out the elements that satisfy the more restrictive
definition.  Thus the starting point of the discussion of algorithmic
decidability of n-manifolds and n-conifolds is to find an appropriate
algorithmically decidable set that includes these spaces.

\subhead{\undertext{4.2~Algorithmic Decidability of Manifolds and
Conifolds}}

For simplicity it is useful to restrict the discussion of algorithmic
decidability to  closed connected manifolds and
conifolds.   The results are easily extended to the case of compact
connected spaces with boundary by doubling the complex over at the
boundary and applying the closed results.\refto{caveat} The set of
complexes $K$ containing a finite number of simplices is clearly an
algorithmically decidable set as one simply checks that the list of
simplices in $K$ and the set of rules for constructing the topological
space $|K|$ satisfy Def.(2.5).  The algorithm must be finite as the
number of simplices is finite.  However, for reasons of efficiency it
is better to use a smaller algorithmically decidable set as the
starting point, the set of all closed n-pseudomanifolds.  It is easy to
see that this set is  algorithmically decidable: First observe that all
complexes in this set contain a finite number of simplices as they are
closed and connected. Therefore, begin with the decidable set of
complexes containing a finite number of simplices.  Next, given a
complex $K$ in this set, find
the maximum dimension n of any simplex in $K$ and then check that all
simplices of dimension less than n are contained in an n-simplex to
verify that the complex is pure. As the the number of simplices to be
checked is finite, the algorithm to do
this check is finite. Second, verify that the simplicial complex is
nonbranching, that is each (n-1)-simplex is contained in precisely two
n-simplices. This can be readily done in terms of the rules used for
constructing the complex and again is a finite procedure.
Finally, verify condition {\sl i)} of Def.(2.8). Again this procedure
can be readily done and is finite:
\proclaim Procedure (4.1).  Observe that condition {\sl i)} is
transitive; if simplex A is connected to simplex B by an appropriate
sequence of n-simplices and if B is connected to C by another
appropriate sequence, then A is connected to C by the sequence
consisting of the concatenation of these two sequences. Next
 simply begin with an arbitrary n-simplex A and construct the set
${\cal A}^1$ consisting of A and all n-simplices that adjoin A by a
(n-1)-simplex. Then  construct  the set ${\cal A}^2$ consisting of
${\cal A}^1$ and all simplices that adjoin ${\cal A}^1$ by a
(n-1)-simplex.  Repeat this procedure $N$ times where $N$ is the number
of n-simplices in the complex  to construct ${\cal A}^N$. Either ${\cal
A}^N$ contains all n-simplices in the complex or it does not; simply
counting the number of n-simplices in ${\cal A}^N$ will determine
this.\par

\noindent Therefore this finite procedure determines whether or not the
requirement is satisfied. These three finite algorithms determine
whether or not $K$ is a closed connected pseudomanifold and therefore
the set of closed connected n-pseudomanifolds is algorithmically
decidable.

In one dimension, all closed combinatorial 1-conifolds are 1-manifolds
and all closed connected 1-manifolds are 1-pseudomanifolds by
definition.  Therefore, immediately from the above result, closed
connected 1-manifolds are algorithmically decidable.  Alternately, this
can be seen explicitly; let $P^1$ be a closed 1-pseudomanifold.
Observe that there are a minimum of three 1-simplices in $P^1$ because
the vertices of a simplicial complex completely determine the complex.
Next pick any vertex $v$ in $P^1$. As $P^1$ is a closed pseudomanifold,
there are exactly two 1-simplices which meet at $v$. Thus $St(v)$ is a
line segment. Therefore the neighborhood of every vertex manifestly
satisfies Def.(2.14). There are precisely two 1-manifolds; the circle
$S^1$ which is a closed manifold and the real line which is not a
closed manifold. As the $P^1$ is closed, it must be equivalent to
$S^1$. Thus, given a space that is a closed connected 1-pseudomanifold,
it is a closed connected 1-manifold and no additional algorithm is
needed to differentiate it.

In two or more dimensions, closed n-pseudomanifolds are more general
than closed connected combinatorial n-manifolds and n-conifolds.
However, these spaces are subsets of the set of closed
n-pseudomanifolds. Thus a first step to an algorithmic description of
these  combinatorial spaces is to have a description of how they differ
from pseudomanifolds. A useful tool is the following result.

\proclaim Theorem {(4.2)}. Let $v$ be any vertex of a closed
n-pseudomanifold $P^n$. Then the link $L(v)$ is a pure nonbranching
(n-1)-simplicial complex without boundary.

As $X^n$ is pure, the star $St(v)$ of each vertex $v$ is a pure
simplicial complex of dimension n.  Each n-simplex in $St(v)$ is
uniquely specified by (n+1) vertices, one of which is $v$ and thus each
n-simplex in $St(v)$ has precisely one (n-1)-simplex  $\sigma^{n-1}_L$
that does not contain $v$. It also follows that this (n-1)-simplex is
uniquely specified by n vertices.
Additionally, all lower dimensional simplices in the n-simplex that do
not contain $v$ are subsets of set of these n vertices uniquely
specifying  $\sigma^{n-1}_L$.
Consequently, each n-simplex in $St(v)$ contributes to $L(v)$ the
complex consisting of this one (n-1)-simplex $\sigma^{n-1}_L$ and its
lower dimensional faces.
(For example, consider the star of the vertex $i$ in Figure 6b) of the
suspension of $RP^2$. Tetrahedra $icbg$ contributes triangle $cgb$ and
its faces to $L(i)$.) Therefore, $L(v)$ is a pure (n-1)-complex.

In order to prove that $L(v)$ is nonbranching, it must be shown that
each (n-2)-simplex in $L(v)$ is contained in precisely two
(n-1)-simplices.
Note that each (n-2)-simplex that is in $L(v)$ is also a face of a
(n-1)-simplex containing $v$. As $X^n$ is nonbranching, each
(n-1)-simplex in $St(v)$ that contains the vertex $v$ is in exactly two
n-simplices. (For example, edge $bg$ is in triangle $bgi$  in Figure
6b) and triangle $bgi$ is in tetrahedra $bgci$ and $bgfi$). It follows
that the two (n-1)-simplices in $L(v)$ that come from these adjoining
n-simplices share a common (n-2)-simplex in $L(v)$. (For example, edge
$bg$ is in triangles $cbg$ and $fbg$.) Thus each (n-2)-simplex is in at
least two (n-1)-simplices in $L(v)$. Next, if some (n-2)-simplex in
$L(v)$ were to be contained in more than two (n-1)-simplices in $L(v)$,
it would have to be a (n-2)-simplex in more than two n-simplices in
$St(v)$.  However, this would imply that $X^n$ is branching; the (n-1)
vertices of the (n-2)-simplex and $v$ uniquely specify a (n-1)-simplex
containing $v$ and thus the (n-1)-simplex containing $v$ would be a
face of more than two n-simplices. Consequently each (n-2)-simplex is
contained in precisely two (n-1)-simplices in $L(v)$. Therefore $L(v)$
is a pure, nonbranching simplicial complex. QED.

This theorem will be used as a tool in finding algorithms for deciding
whether or not a closed  n-pseudomanifold is a closed connected
combinatorial n-manifold or n-conifold in low dimensions. The case of
closed connected n-manifolds will be discussed first.  In order to
decide whether or not a given n-pseudomanifold is actually a closed
connected combinatorial n-manifold by Def.(2.14), one needs a finite
algorithm for deciding whether or not the link of every vertex of the
n-pseudomanifold is actually a combinatorial \hbox{(n-1)-sphere}.
Whether or not this can be done is dependent on the dimension.  It
turns out that closed connected combinatorial n-manifolds are
algorithmically decidable in two and three dimensions,\refto{stillwell}
and can be proven to be not decidable in five or more
dimensions.\refto{rs,freedman} Whether or not 4-manifolds are decidable
is an open question.

In order to show that 2-manifolds are algorithmically decidable,
consider how they differ from 2-pseudomanifolds. Let $P^2$ be a closed
connected 2-pseudomanifold.  First note that the link of every vertex
of $P^2$ is the disjoint union of circles. In order to prove this, let
$v$ be any vertex of $P^2$. The link $L(v)$  is a pure nonbranching
1-dimensional simplicial complex without boundary by Thm.(4.2). Note
that $L(v)$ may have several disconnected components. However, each
component must be a 1-pseudomanifold because condition {\sl i)} in
Def.(2.8) is equivalent to being connected in one dimension. Therefore,
$L(v)$ is the disjoint union of closed 1-pseudomanifolds, or
equivalently the disjoint union of circles. Thus if $L(v)$ is connected
for every vertex $v$ of $P^2$, then $P^2$ is actually a 2-manifold.
Furthermore, the character of the links
implies any closed 2-pseudomanifold that is not a 2-manifold is
obtained by identifying vertices of a closed connected combinatorial
2-manifold.

Given these properties, it is easy to find an algorithm to determine
whether or not a given 2-pseudomanifold $P^2$ is actually a 2-manifold.
First find all the links of the \hbox{2-pseudomanifold}. Then check
that every link is connected;  this can be done using the procedure
(4.1) in this one dimensional case. These two steps are manifestly both
finite procedures; thus they constitute a finite algorithm for deciding
2-manifolds. Therefore closed connected 2-manifolds are decidable.

In order to show that closed connected 3-manifolds are algorithmically
decidable, again consider how they differ from closed
3-pseudomanifolds. Let $P^3$ be a closed connected 3-pseudomanifold and
$v$ be any vertex in $P^3$. The link $L(v)$ is a pure nonbranching
2-complex. Therefore, in order to test whether or not $P^3$ is a
3-manifold one needs an algorithm to test for whether or not a pure
nonbranching 2-complex is a 2-sphere.  This can be done in two finite
steps: First  check whether or not $L(v)$ is a closed 2-pseudomanifold.
This can be done by applying the finite algorithm for deciding
2-pseudomanifolds discussed previously. Next, if $L(v)$ is found to be
a 2-pseudomanifold, whether or not it is a 2-sphere can be determined
by computing its Euler characteristic.  The Euler characteristic of any
pure  n-complex
is given by the alternating sum $$\chi(K^n) = \sum_{i=0}^n
(-1)^{i}\alpha_i\eqno(euler)$$ where $\alpha_i$ is the number of
i-simplices in the complex. The Euler characteristic of a 2-sphere is
two and it is a well known fact that the Euler characteristic of any
other closed 2-manifold is less than two. In addition it is easy to
prove that the Euler characteristic of any closed 2-pseudomanifold is
less than or equal to two by using the previously mentioned fact that
all 2-pseudomanifolds are obtained from 2-manifolds by the
identification of vertices. Each identification on a 2-manifold lowers
the Euler characteristic of the resulting 2-pseudomanifold by one;
therefore the Euler characteristic of a 2-pseudomanifold is two if and
only if it is a 2-sphere. Thus there is a finite algorithm for
determining whether or not $L(v)$ is a 2-sphere.

Given these properties, the algorithm for determining whether or not
$P^3$ is a 3-manifold follows directly. First find all the links of
$P^3$. Next check that every link is a 2-sphere by the algorithm
described above. If all links are found to be 2-spheres by this
procedure then $P^3$ is a 3-manifold. This algorithm is manifestly
finite and thus closed connected 3-manifolds are decidable.

However in dimensions higher than three, one runs into difficulty.
Recall that the link of a vertex being a combinatorial (n-1)-sphere  is
equivalent to its star being a combinatorial n-ball. Moreover, one can
show that there is no algorithm for recognizing  a n-ball for $n\ge
6$.\refto{rs} This proof has become part of mathematics folklore; it is
originally due to S. Novikov but has never been published. Furthermore,
using modern results one can extend this proof to demonstrate that
there is no algorithm for recognizing a 5-ball, as will be outlined
below.\refto{freedman} It is a topological, not explicitly simplicial
proof but obviously applies to the simplicial case as well. The proof
utilizes the h-cobordism theorem:\refto{rs, freedman} Given any compact
(n+1)-manifold $Y^{n+1}$ with $n\ge 4$ and two boundary components
$M_1^n$ and $M_2^n$ such that
$\pi_1(M_1^n)=\pi_1(M_2^n)=\pi_1(Y^{n+1})=1$ and $H_*(Y^{n+1},
M_1^n)=0$, then $Y^{n+1}$ is homeomorphic to $M_1^n\times I$.  Next,
note that one can construct a special type of closed n-manifold $M^n$
with $n\ge 5$ that has any finitely presented group as its fundamental
group. This part of the construction is given in the next subsection as
it is also directly relevant to the classifiability of n-manifolds.
Next, let $W^n$ be $M^n$ minus the interiors of two disjoint n-balls;
thus $W^n$ has boundary consisting of two disjoint n-spheres and  has
the same fundamental group as $M^n$. Furthermore, one can show that
there is an infinite set of groups so that $\pi _1(W^n)=1$ implies that
$H_*(W^n,S^{n-1})=0$. Thus $W^n$ satisfies the conditions of the
h-cobordism theorem if $\pi_1(W^n) =1$.  In particular, the above
manifold $W^n$ for $n\ge 5$ is a product $S^{n-1}\times I$ if and only
if its fundamental group is trivial. Therefore, by capping off one of
the boundary spheres of $W^n$ to form the manifold ${\cal B}^n$, it
follows that ${\cal B}^n$ is a n-ball if and only if its fundamental
group is trivial.  However, one can prove that there are groups for
which the word problem is unsolvable in the set of fundamental groups
satisfying the conditions of the h-cobordism theorem. Thus there is no
algorithm to show that every word in $\pi_1({\cal B}^n)$ is trivial for
all possible groups. The conclusion that there is no algorithm follows
directly; ${\cal B}^n$ is homeomorphic to an n-ball if and only if its
fundamental group is trivial but no algorithm exists for determining
this fact. Therefore, there is no algorithm to decide if a given space
is a n-manifold for $n\ge 5$ because there is no way to decide if the
neighborhoods of the space are equivalent to n-balls.

The algorithmic decidability of 4-manifolds is an open problem.  An
algorithm for recognizing a combinatorial 4-manifold requires  an
algorithm for recognizing a combinatorial 3-sphere.  There are two
sufficient conditions for the existence of such an algorithm: a
solution to the Poincar\'e conjecture, and a solution to the word
problem for the fundamental groups of 3-manifolds, both open problems
in topology.\refto{stillwell} The Poincar\'e conjecture is that any
closed simply connected 3-manifold is actually a 3-sphere. If the
Poincar\'e conjecture is true, then it provides a starting point for
developing an algorithm for recognizing a 3-sphere. The next step is to
find a computable method of determining whether or not the fundamental
group of a 3-manifold is trivial.  Whether or not such an algorithm
exists is also an open issue; one can prove that the set of all
fundamental groups of 3-manifolds is only a subset of the set of all
finitely presented groups. Moreover, it is currently unknown as to
whether or not the word problem is solvable for this subset of finitely
presented groups.\refto{stillwell} Therefore,
if the Poincar\'e conjecture holds and the word problem for the
fundamental groups of 3-manifolds has an algorithmic solution, then
there will exist  a finite algorithm for recognizing a combinatorial
3-sphere. If the Poincar\'e conjecture is not true, it may still be
possible that a finite algorithm exists for recognizing a 3-sphere. For
example, if the word problem for the fundamental groups of 3-manifolds
is solvable and if a known set of counterexamples to the Poincar\'e
conjecture that are all characterized by a finite algorithm is found,
then a finite algorithm for recognizing the 3-sphere could also be
constructed by combining these results. Alternatively, there may be
some method of recognizing a 3-sphere that does not rely on either the
Poincar\'e conjecture or the word problem for the fundamental groups of
3-manifolds.  In any case, it does not  seem likely that the algorithm
for recognizing a combinatorial 3-sphere will be simple if any does
exist; after all, mathematicians have tried to find one for nearly 100
years. Without such an algorithm, it is not possible to decide whether
or not a given 4-pseudomanifold $P^4$ is a 4-manifold. Thus there is
currently no known finite algorithm for algorithmically deciding
4-manifolds.

The algorithmic decidability of n-conifolds is studied in exactly the
same manner as that for n-manifolds.  By Def.(2.16), conifolds in one
and two dimensions are manifolds; thus it follows from the previous
results that closed connected conifolds are algorithmically decidable
in one and two dimensions. In three or more dimensions, conifolds are
more general topological spaces than manifolds and therefore the
algorithmic decidability of conifolds in these dimensions is a separate
issue. In the case of n-conifolds, one needs a finite algorithm for
determining whether or not the link of every vertex is a combinatorial
\hbox{(n-1)-manifold}. Again, as in the manifold case, whether or not
this can be done is dependent on dimension. It turns out that closed
connected n-conifolds are algorithmically decidable not only in three
but also in four dimensions. Additionally it can be proven that closed
connected n-conifolds are not decidable in six or more dimensions and
the question remains open in five.

In order to show that closed connected 3-conifolds are decidable, again
begin by considering how they differ from closed 3-pseudomanifolds.
Let $v$ be a vertex of a closed 3-pseudomanifold $P^3$. By Thm.(4.2),
the link $L(v)$ is a pure nonbranching  simplicial 2-complex without
boundary.  Therefore, in order to test whether or not $P^3$ is a
3-conifold, one needs an algorithm to test for whether or not a pure
nonbranching 2-complex is a 2-manifold. This can be done in two finite
steps: First  check whether or not $L(v)$ is a closed 2-pseudomanifold.
This can be done by applying the finite algorithm for deciding
2-pseudomanifolds discussed previously. Next, if $L(v)$ is indeed a
2-pseudomanifold, apply the finite algorithm for deciding whether or
not a 2-pseudomanifold is  a closed 2-manifold.
Therefore, given the algorithm for determining whether or not $L(v)$
is a closed 2-manifold, the algorithm for determining whether or not
$P^3$ is a 3-conifold is to repeat this procedure for each vertex of
$P^3$. If each link $L(v)$ is a closed 2-manifold, then $P^3$ is a
closed combinatorial 3-conifold. Thus 3-conifolds are algorithmically
decidable.

Given that all 3-manifolds are 3-conifolds, it is useful to have an
algorithm for determining whether or not a given closed 3-conifold
$X^3$ is actually a combinatorial 3-manifold. First,   note that all
triangulations of 3-manifolds are combinatorial triangulations as
observed in section 3.  Next, recall from Thm.(5.2) of part I that a
3-conifold is a 3-manifold if and only if its Euler characteristic is
zero. Thus calculate the Euler characteristic of $X^3$ using \(euler);
this procedure is obviously finite.  It follows that if $\chi(X^3) =
0$, then $X^3$  is a combinatorial 3-manifold. Thus the test for
whether or not a given combinatorial 3-conifold is actually a
combinatorial 3-manifold is very simple.

In order to show that closed connected 4-conifolds are decidable, again
consider how they differ from 4-pseudomanifolds. Let $P^4$ be a closed
4-pseudomanifold. The link $L(v)$ of each vertex $v$ is a pure
nonbranching simplicial 3-complex without boundary and thus, in order
to proceed, an algorithm is needed to determine whether or not a pure
nonbranching 3-complex is a closed connected 3-manifold. Such an
algorithm is the following: First test whether or not $L(v)$ is a
3-pseudomanifold by applying the finite algorithm for deciding
3-pseudomanifolds. If $L(v)$ is indeed a 3-pseudomanifold, then test
whether or not the Euler characteristic of $L(v)$ is zero; if the Euler
characteristic is nonvanishing then $L(v)$ cannot be a 3-manifold. If
the Euler characteristic indeed vanishes, test that $L(v)$ is actually
a closed connected combinatorial 3-manifold by applying either the
finite algorithm for deciding 3-manifolds or the finite algorithm for
deciding 3-conifolds.
This is a finite algorithm for deciding when $L(v)$ satisfies
Def.(2.14).
{}From it one deduces that the finite algorithm for deciding when $P^4$
is
combinatorial 4-conifold is to repeat this algorithm for every vertex
$v$ in $P^4$. If each link is found to be a closed connected 3-manifold
then $P^4$ is a combinatorial 4-conifold; otherwise it is not. Thus,
4-conifolds are algorithmically decidable.

In five dimensions, whether or not the class of 5-conifolds is
algorithmically decidable is open; it depends on whether or not
4-manifolds are algorithmically decidable. Obviously,  if closed
connected combinatorial 4-manifolds are eventually found to be
algorithmically decidable, then 5-conifolds will also be
algorithmically decidable and conversely. Finally, it is easy to prove
that n-conifolds are not decidable in  six or more dimensions by
observing that n-manifolds are not decidable in five or more
dimensions. Therefore there is no algorithm for recognizing when the
link of a vertex in a n-pseudomanifold is actually a combinatorial
(n-1)-manifold for $n\ge6$. Thus combinatorial n-conifolds in six or
more dimensions are not algorithmically decidable.

Finally, it should be emphasized that there is no method of
constructing a set of topological spaces that are not proven to be
algorithmically decidable contrary to a certain suggestion in the
literature by Hartle.\refto{jbh1} Hartle asserts that a set of all
4-manifolds can be constructed in terms of combinations of a set of
known combinatorial manifolds. Such an assertion may seem reasonable
given certain theorems in surgery of n-manifolds as described
below.\refto{rs, freedman} Given the set of all closed n-manifolds, one
can define an equivalence relation by defining two to be equivalent if
and only if they are cobordant.  Applying this equivalence to the set
of all closed n-manifolds yields a finite set consisting of the
equivalence classes in dimension n. One can now pick a known explicit
manifold to represent each equivalence class. These manifolds can be
taken to be the finite set of building blocks for generating all
n-manifolds. The surgical result used to do this
 is that two n-manifolds are cobordant if and only if they differ by a
finite number of handle surgeries.\refto{surgery}  Namely, embed
$S^p\times B^{q+1}$ with $p+q+1=n$ in a n-manifold $M^n$, remove its
interior yielding a compact n-manifold with boundary $S^p\times S^q$,
then glue $B^{p+1}\times S^q$ in along the boundary $S^p\times S^q$ to
yield a closed n-manifold. In two dimensions, this procedure is simply
that of adding handles to $S^2$ and $RP^2$ to generate all 2-manifolds.
Thus this is just the natural higher dimensional generalization.
However note that surgery techniques can  be used to construct all
n-manifolds and  independently, it is known that n-manifolds are not
algorithmically decidable in five or more dimensions as outlined above
for the particular case of the n-ball.  The problem is that in order to
algorithmically implement such surgery techniques, one must prove that
they can be encoded in an explicit finite algorithm. This clearly can
be done in two and three dimensions. However, it turns out that the
different embeddings of the handles are not nice in higher dimensions
and it is this step that is not algorithmically describable.  Thus
Hartle's scheme for building the set of all 4-manifolds cannot be
proven to actually generate this set without an algorithm for
recognizing the 3-sphere.
Therefore the issue of the algorithmic decidability of 4-manifolds
cannot be avoided.

\subhead{\undertext{4.3~Classifiability of Manifolds and
Conifolds}}

In order to discuss the classifiability of manifolds and conifolds, it
is first necessary to state the criteria by which two combinatorial
manifolds or conifolds will be judged to be physically  distinct.  As
just seen in the algorithmic decidability subsection, the criteria used
for defining a set is strongly coupled to the existence of an algorithm
for implementing it. The standard criteria in the mathematics
literature for classifying n-manifolds or n-conifolds in the continuum
is equivalence under homeomorphisms; two topological spaces are said to
be equivalent if they are homeomorphic to each other. However, in more
than four dimensions, two spaces can be homeomorphic but not
diffeomorphic as they can have distinct smooth structures. Moreover,
the physically natural invariance applied to histories appearing in
expressions \(topchange) or \(fpartfn) is diffeomorphism invariance.
Thus the desired criteria for physics is to classify two smooth
topological spaces as physically equivalent if they are both
homeomorphic to each other and have equivalent smooth structures. In
less than seven dimensions, this criteria is equivalent to the
combinatorial equivalence of the corresponding combinatorial
counterparts of the smooth topological spaces. Thus the natural
criteria for the classification of combinatorial n-manifolds and
n-conifolds is combinatorial equivalence.  Note that by the results of
section 3, a sum over combinatorially inequivalent topological spaces
incorporates a sum over smooth structures.  Thus combinatorial
equivalence takes care of both the issue of equivalence under
homeomorphisms and equivalence of smooth structures in a very natural
fashion.

Again the classification of n-manifolds and n-conifolds will be
discussed
for the case of closed connected combinatorial manifolds and
conifolds; it is easy to prove that the results can be directly
extended to compact spaces. As all conifolds are manifolds in
dimensions one and two, it follows immediately that they can be
classified if 1-manifolds and 2-manifolds can be classified. Closed
connected 1-manifolds are obviously classifiable; the only closed
connected 1-manifold is a circle. Therefore the set of distinct
1-manifolds has only one element. Closed connected 2-manifolds are
classified by the orientability or nonorientability of the manifold and
its Euler characteristic.\refto{stillwell} Given a closed combinatorial
2-manifold $M^2$, its orientability or nonorientability can be
determined  by computing its second homology; $H_2(M^2) = {\bf Z}$ if
it is orientable and $H_2(M^2) = 0$ if it is not. This computation
clearly takes a finite number of steps as the number of simplices in
the closed combinatorial 2-manifold is finite.
The Euler characteristic \(euler) of the 2-manifold is also clearly
calculated in a finite number of steps.  Thus these two procedures form
a finite algorithm and a set of distinct 2-manifolds can be generated
using it. As the Euler characteristic is unbounded below, this set
contains a countably infinite number of representatives.

Whether or not combinatorial 3-manifolds or 3-conifolds are
classifiable is an open problem.\refto{stillwell} Clearly, a necessary
step in solving this problem is an algorithmic method of recognizing
combinatorial 3-spheres. As discussed extensively in the algorithmic
decidability subsection,  sufficient conditions for solving this
problem are an algorithmic solution to the Poincar\'e conjecture and to
the word problem for fundamental groups of 3-manifolds.
 Given an affirmative solution to the problem of algorithmically
recognizing a 3-sphere, it may be possible to provide a finite
algorithm for classifying closed connected 3-manifolds if other open
conjectures for 3-manifolds can also be solved. However, the issue is
clearly unresolved at the present time.  Similarly it turns out that
the classifiability of 3-conifolds is also an open problem. It follows
from the observation that all finite presentations of groups do not
appear as the fundamental groups of 3-conifolds that there is no easy
proof that 3-conifolds are not classifiable. Then the observation that
the problem of classifying 3-manifolds is open immediately implies that
the issue is also open for 3-conifolds as 3-manifolds are a proper
subset of 3-conifolds.  Thus there is no known method of generating a
sets of either combinatorially inequivalent 3-manifolds or
3-conifolds.

Independently of the issue of whether or not  the set of closed
connected combinatorial \hbox{4-manifolds} is algorithmically
decidable, it can be proven that they are
 not classifiable.\refto{4class} This is a consequence of the following
topological argument that all finitely presented groups occur as the
fundamental group of some n-manifold for $n\ge 4$; although presented
in general terms, it is clear that it can be implemented in terms of
combinatorial triangulations of these n-manifolds and thus applies in
the simplicial case.  Given any finitely presented group $G$ and some
fixed $n\ge 4$, there is a closed smooth n-manifold with $G$ as its
fundamental group. First, one can produce a n-manifold $M^n$ with
fundamental group of $k$ generators and no relations by taking the
connected sum of $k$ copies of the n-manifold $S^1\times S^{n-1}$ with
itself. Next, observe that if a simple closed curve $c$ is removed from
any manifold of four or more dimensions,  the fundamental group remains
the same.  Similarly, if a tubular neighborhood around that curve is
removed, $\pi_1(M^n)$ also remains the same.\refto{3dbreakdown} The
tubular neighborhood is $S^1\times B^{n-1}$ and its boundary is
$S^1\times S^{n-2}$.  Now, $S^1\times S^{n-2}$ is also the boundary of
$B^2\times S^{n-2}$.  Hence, one can replace the interior of the
tubular neighborhood in $M^n$ with $B^2\times S^{n-2}$ and obtain a new
manifold $M^n_c$. Now, it is easy to check that the curve $c$ is now
contractible to a point in $M^n_c$. Hence by taking the curve $c$ to
pass through the appropriate copies of $S^1\times S^{n-1}$ in the
appropriate order, this construction gives one relation $c=1$ among the
$k$ generators.  Thus $\pi_1(M^n_c)$ is a group with $k$ generators
and  one relation. By repeating this surgery procedure a finite number
of times on other curves, any set of relations for a group with $k$
generators can be derived. Therefore, any finitely presented group is
the fundamental group of some smooth closed n-manifold for $n\ge 4$.

Furthermore, one can prove that a subset of the set of all closed
n-manifolds can be constructed such that their equivalence is
completely determined by their fundamental groups; that is two
n-manifolds in this subset are homeomorphic if and only if their
fundamental groups are isomorphic. However, the isomorphism problem for
finitely presented groups is unsolvable; immediately this implies that
there is no algorithm for deciding when two n-manifolds in this subset
are homeomorphic.  Consequently, by the observation that a set is not
classifiable if any subset of it is not, there can be no finite
algorithm for classifying the set of all smooth closed n-manifolds.
This result also clearly applies to the combinatorial equivalence of
n-manifolds as well. Finally,  as the set of all closed connected
combinatorial n-conifolds contains all closed n-manifolds, it
immediately follows that all n-conifolds in dimensions four or more
cannot be classified by the above arguments.  Therefore, one cannot
classify either closed combinatorial n-manifolds  or n-conifolds of
dimension $n\ge 4$.

It should be emphasized that such n-manifolds as described above can
be explicitly constructed from  known  examples of finitely presented
groups that have no solution to the word problem. In particular, a
manifold with the fundamental group given in Appendix C can be
explicitly constructed. This
manifold can be combinatorially triangulated and thus it would appear
in any simplicial approximation to manifolds using a sufficient
number of n-simplices. Thus the fact that n-manifolds are not
classifiable in four or more dimensions is not simply a difficulty in
principle, but a problem that can and will actually be encountered even
in the construction of a set of spaces that contain a finite number of
n-simplices.

Note that the issue of classifying other topological spaces besides
manifolds and conifolds can be studied by similar methods. For example,
2-pseudomanifolds are algorithmically classifiable in two dimensions.
In fact 2-complexes themselves are algorithmically classifiable;
however interestingly enough it is not possible to decide whether or
not $\pi_1(K^2)$ of a 2-complex is trivial.\refto{greek} This is
because the equivalence of 2-complexes under homeomorphisms
is not in one to one correspondence with the isomorphism problem for
their fundamental groups.  The problem of classifying 3-pseudomanifolds
is open and in four or more dimensions, n-pseudomanifolds are known to
be not classifiable from the observation that n-manifolds are a subset
of n-pseudomanifolds and are not classifiable.

Finally, it is very important to take care when applying results quoted
in the mathematics literature on classifiability to the issue of
finding a set of physically distinct spaces.  As emphasized in this
section, finding a set of physically distinct spaces involves having
algorithms for  deciding when a space is a member of the desired set
and for deciding when it is distinct.  However, it is common in the
mathematics literature to not require that there be such algorithms for
a set to be called classifiable. For example, a well known statement is
that $K(\pi,1)$ manifolds are classifiable;\refto{kpi1} there is a well
known theorem that states that two $K(\pi,1)$ manifolds are homotopy
equivalent if and only if their fundamental groups are isomorphic.
However, by the isomorphism problem for groups,
there is no algorithm for determining whether or not two arbitrary
fundamental groups are isomorphic. Therefore, although $K(\pi,1)$
manifolds are classifiable, there is no algorithm for doing so.
Similarly, a particular example  has been frequently misunderstood in
physics literature is the case of simply connected spin 4-manifolds. It
is a well known fact that simply connected spin 4-manifolds are
classified by their Euler characteristic and signature.\refto{freedman}
However, again by the word problem for groups, there is no algorithm
for determining whether or not an arbitrary 4-manifold is simply
connected as
one can not prove that any arbitrary closed curve is contractible to
a point.  It follows that simply connected spin 4-manifolds cannot be
algorithmically classified simply because the set of simply connected
spin 4-manifolds cannot be algorithmically decided. Consequently, a sum
over these spaces can not be concretely implemented contrary to what is
frequently stated in the physics literature.\refto{sh,jbh1} Thus,
one must be careful to check whether or not a classifiable set is
algorithmically classifiable before claiming a sum can be formulated in
terms of it.

\head{5.~Euclidean Functional Integrals using Regge Calculus}

The results of the last section provide the necessary background for
 explicitly implementing  sums over topological spaces. Although the
previous discussion is directly applicable to both the continuum and
discrete formulations of sums over histories,  it is useful to
illustrate the  explicit implementation of such sums  in terms of
simplicial complexes.  The resulting sums over histories provide a
discrete approximation of quantum amplitudes  can be implemented
numerically and thus
a quantitative study of the consequences of a sum over topology can be
 directly carried out. Moreover, the effects of different algorithms
for generating the necessary lists of spaces are also readily
accessible for amplitudes constructed completely in terms of simplicial
complexes. Therefore, this section will first discuss Regge calculus
and then discuss  algorithms for summing over manifolds and conifolds
with emphasis on the case of four dimensions.  As in section 4, it is
easiest to address the topological aspects of implementing sums over
histories in terms of closed connected manifolds or  conifolds such as
\(fpartfn). However, with more work, these results can be readily
applied to sums over histories involving compact spaces such as
\(topchange) as well.

In order to translate an expression such as \(fpartfn) into a concrete
sum over simplicial histories it is necessary to be able to associate a
metric and action with any $K^n$. To this point in the paper, no metric
information has been associated with the simplicial complexes; the
simplicial complexes carry only the topology and PL structure of the
spaces. Thus
 in order to proceed, metric information must be attached. The easiest
 way to do so, as discussed by Regge,\refto{regge} is to require that
the metric on the interior of each n-simplex is the Euclidean metric;
that is all n-simplices in the simplicial complex are flat. With this
requirement, the geometry of each n-simplex is completely fixed by
specifying the lengths  $s_i$ of all of its edges. It follows that the
geometry of the simplicial complex is also completely fixed by
specifying the length of all edges $s_i$ in the complex. Therefore one
anticipates that all geometrical quantities such as volume and
curvature can be expressed completely in terms of the edge lengths.

Indeed this is the case. It is easy to see that volume of any pure
simplicial complex can be computed by first computing the volume of
each n-simplex in terms of the edge lengths and then adding the
contribution of all the n-simplexes.  Somewhat less obviously,
curvature can also be expressed in terms of the edge lengths. As the
metric on the interior of each n-simplex is flat, it is clear that the
curvature of the combinatorial space is not carried on the interiors of
the n-simplices. Rather, it turns out to be concentrated on the
(n-2)-simplices of the simplicial complex.   This is most directly
apparent in two dimensions, in which curvature is concentrated on
vertices. For example, let $ v$ be a vertex in some combinatorial
closed 2-manifold and let $m$ denote the number of triangles in
$St(v)$.
Then the scalar curvature associated with this vertex is given by
$\theta(v)$ where
$$\theta(v) = 2\pi - \sum_{i=1}^{m}\phi_{ i}\eqno(2regcurv)$$
and $\phi_i$ is simply the angle between two unit vectors that lie in
adjacent edges of  the ith triangle  as illustrated in Figure 7a).
 Indeed the sum of the curvature at each vertex over all vertices in
the 2-manifold yields the Euler characteristic, ${\displaystyle
\sum_{v\in M^2}} \theta(v) = 2\pi \chi(M^2)$ as required by the
Gauss-Bonnet theorem.  In general, if the index $i = 1\ldots m$
sequentially labels adjacent n-simplices in $St(\sigma^{n-2})$, then
the curvature is given by
$$\theta(\sigma^{n-2}) = 2\pi -
 \sum_{i=1}^{m}\phi_i\eqno(regcurv)$$
where $\phi_{i}$ is now the
dihedral angle between two unit vectors normal to $\sigma^{n-2}$ that
lie in the adjacent (n-1)-simplices of the ith n-simplex as illustrated
in the three dimensional case in Figure 7b).  All dihedral angles can
be computed in terms of the edge lengths by elementary trigonometry in
any dimension. Therefore, the curvature can be expressed as a function
of edge lengths alone.  In addition, one can demonstrate that the
definition of curvature given in \(regcurv) converges to the scalar
curvature in the continuum in a suitably defined average in greater
than two dimensions.\refto{cheeger} Thus the Regge curvature is a
suitable discrete version of curvature for use in simplicial gravity.

Note that the formulation of the  Regge curvature implicitly relies on
the combinatorial nature of the  simplicial complex. The definition of
the geometry in terms of the edge lengths relies on the fact that the
n-simplices are completely defined by their vertices. Similarly the
definition of curvature relies on the fact that there is a method of
sensibly associating n-simplices and (n-1)-simplices to a given
(n-2)-simplex.  In order to do this,  certain types of simplicial
complexes must be excluded; examples of spaces which do not have the
necessary notions are branching simplicial complexes. In general, one
requires that the simplicial complex be pure and  nonbranching such
that the concept of adjacent (n-1)-simplices is well defined. As
manifolds, conifolds and even pseudomanifolds all satisfy this
condition by definition, Regge calculus can be used to compute
curvatures on all of these topological spaces.

Finally, given the above definitions, the Regge action for Einstein
gravity with cosmological constant $\Lambda$ for a closed pure
nonbranching complex $K^n$ is
$$I[s_i]= -\frac {2}{16\pi G}\sum_{\sigma^{n-2} \in K^n}
\theta(\sigma^{n-2} )V(\sigma^{n-2} )  + \frac { 2\Lambda} {16\pi
G}\sum_{\sigma^{n} \in K^n} V(\sigma^{n})\eqno(raction)$$
where the first sum is over all (n-2)-simplices in the simplicial
complex and the second is over all n-simplices in the complex.
$V(\sigma^n)$ is the volume of the indicated n-simplex. The Regge
action for  compact combinatorial manifolds and conifolds
with boundary  can also be formulated entirely in terms of edge
lengths;\refto{harsor} essentially one adds the appropriate discretized
form of the boundary term that appears in the continuum action
\(topchange). Finally, note that discretized actions for other theories
such as curvature squared theories on simplicial complexes can also be
formulated in terms of the edge lengths.\refto{hamberwilliams}

Given a method of associating a geometry and an action with any complex
$K^n$ in the sets of interest, the Regge  equivalent of \(fpartfn) is
$$\eqalignno{<A> &= \frac {\displaystyle \sum_{K^n \in \cal L}
<A>_{_{K^n}} }{\displaystyle \sum_{K^n \in \cal L} <1>_{_{K^n}}}  \cr
 <A>_{_{K^n}}&=\int Ds\ \ A(s_i)\exp( -I[s_i])&(4simp)\cr}$$
where the sum is over all complexes $K^n$ in an as yet unspecified list
of complexes  $\cal L$. The notation $\int Ds$ denotes the integral
over all edge lengths, $ \int Ds ={\displaystyle \prod_{s_i\in
K^n}}\int d\mu(s_i)$ where the notation $d\mu(s_i)$ indicates the
freedom available in choosing the measure on the space of edge
lengths.  Note that the functional integral over metrics in \(fpartfn)
has been reduced to the product of integrals over edge lengths in
\(4simp) as the metric information is discrete.  This product of
integrals is well defined and thus the sum over all edge lengths is in
principle implementable; discretizing the metric has removed the
problems  related to the issues of gauge fixing and
nonrenormalizability associated with defining the measure on the space
of metrics. (Such problems, of course, reappear when attempting to take
the continuum limit of such a Regge integral.) Moreover, the technical
details of implementing a sum over edge lengths
are manifestly isolated in \(4simp) from those involving the
construction of the list $\cal L$. Thus the issues involved with
algorithmically constructing such a list in different dimensions can be
addressed independently.

Even though the functional integral over metrics has been explicitly
implemented in terms of edge lengths,
the expression \(4simp) is still heuristic; at this point it is
necessary to explicitly provide an algorithm that generates a list of
suitable spaces $K^n$ in the specified set. For example if one wished
to implement \(4simp) as a sum over physically distinct closed
manifolds $M^n$, one would need a
list of combinatorially inequivalent $M^n$. Similarly, an
implementation of \(4simp) as a sum over physically distinct conifolds
necessitates a list of combinatorially inequivalent conifolds.  The
starting point for generating such a list is to generate an exhaustive
list of all topological spaces in the specified set. The second step is
to select from this exhaustive list, the set of unique
representatives.  Thus in order to get anywhere at all, the specified
set of topological spaces must be algorithmically decidable. Next, the
second step requires the space to be classifiable according to the
desired criteria.  Thus, the considerations of  section 4 apply
directly to this issue.

It is useful to begin by discussing the case of closed 2-manifolds, as
it is an explicit example for which both steps can be completely
carried out to form a list of physically distinct 2-manifolds. One can
generate an exhaustive list of 2-manifolds by the following procedure:
Generate all spaces built out of gluing together
n triangles along their edges such that they form simplicial
complexes. Then apply the algorithm for the definition of a closed
2-manifold to this set to find all 2-manifolds made of n triangles.
Finally,  repeat this procedure for all values of n. This procedure is
computable by  the last section. The result will be a list of all
combinatorial 2-manifolds with a vast amount of redundancy.  For
example, one will have a large number of 2-manifolds that are
equivalent under simplicial homeomorphisms, that is under permutations
of the vertices. Additionally one will have a large number of
combinatorially equivalent spaces; for example the list will include
combinatorial 2-spheres composed of four triangles, those of six
triangles, those of 386 triangles and so on, all of which are
combinatorially equivalent.  Thus it is necessary to classify by
combinatorial equivalence this initial list of spaces to eliminate this
redundancy. One can produce such a list of unique representatives of
2-manifolds in the following way: Pick one of the 2-manifolds in the
initial set, say one with the smallest number of triangles, and compute
whether or not it is orientable and compute its  Euler characteristic.
Record its orientability and its Euler characteristic  and place it on
the list of unique representatives $\cal L$.  Next repeat this
computation of orientability and Euler characteristic for each
2-manifold in the initial set of 2-manifolds. For each of these
manifolds, check to see if a 2-manifold with the same Euler
characteristic and orientability already appears on the list $\cal L$
of unique representatives; if it does, discard it and go on to the next
manifold. If it does not, add it to the list. Continue through all
2-manifolds on the initial list. One will end up with a $\cal L$ of all
physically distinct 2-manifolds; the elements in this list are uniquely
specified by their Euler characteristic and orientability.

Note that the unique representatives generated by this procedure will
have different numbers of triangles. It is also clear that there exist
different procedures that will also select out a list of unique
representatives ${\cal L}'$ that contain different combinatorial
2-manifolds as its members; that is the list ${\cal L}'$ will contain a
2-manifold with the same Euler characteristic and orientability as one
in $\cal L$ but this 2-manifold may be
composed of a different number of triangles or differ by a simplicial
homeomorphism. However, these points do not affect the topological
results; any list of unique representatives ${\cal L}$ is
equivalent to any other such list ${\cal L}'$ by the fact that
2-manifolds are classified under combinatorial equivalence by their
orientability and Euler characteristic. Thus the sum over topologies in
\(4simp) can be concretely implemented in terms of any such list of
unique representatives.

Next consider the  construction of such lists in four dimensions for
manifolds and conifolds. The starting point  is the same as that in two
dimensions; an exhaustive list of the set of topological spaces of
interest. It is at this step that problems occur with formulating
expressions for the set of all 4-manifolds; as discussed in section 4,
there is no known algorithm for generating an exhaustive list. Thus,
there is no known starting point for the rest of the algorithmic
formulation of a list.

In comparison, closed 4-conifolds can indeed be algorithmically decided
in  four dimensions and thus the first step can be concretely
implemented. In parallel with the 2-dimensional case, one begins with n
4-simplices, generates all spaces built out of all possible
combinations of these 4-simplices and then applies the algorithm for
the definition of a closed 4-conifold to find all 4-conifolds made of n
4-simplices. One  then repeats this procedure for all values of n. The
resulting exhaustive list $\cal L$ contains all closed 4-conifolds
again with a large amount of redundancy. Thus the set of closed
combinatorial 4-conifolds has an immediate advantage over the set of
all closed combinatorial 4-manifolds; as they are algorithmically
decidable, an exhaustive list of 4-conifolds can be explicitly
constructed.

However, unlike the case in two dimensions, the second step cannot be
carried out to form a list of physically distinct 4-conifolds because
of the results of section 4; neither 4-manifolds nor 4-conifolds are
classifiable under combinatorial equivalence. Therefore a sum over
combinatorially inequivalent 4-conifolds cannot be  implemented even
though an exhaustive list of these spaces can be constructed.
Moreover,  this problem with implementing a sum over 4-conifolds
cannot be removed by requiring them to satisfy certain stricter
criteria, such as requiring them to be simply connected as discussed in
section 4.  Therefore,  as a list of combinatorially inequivalent
4-conifolds cannot be constructed,
\(4simp) cannot be implemented as a sum over these spaces.
Additionally,  the problem with constructing a list of combinatorially
inequivalent 4-dimensional topological spaces cannot be solved for any
algorithmically decidable set that includes all 4-manifolds.

Thus at this point one is forced to conclude that any concrete
formulation of a sum over topology of form \(4simp) in four dimensions
must begin by a reexamination the criteria used in defining what is
meant by a distinct topological space. It is clear that the problems
that arise with the classification of 4-manifolds and 4-conifolds are
closely tied with the criteria used in defining the set of interest.
Thus a natural way to avoid
these problems is to change the criteria for defining what a  distinct
topological space is, that is to use a less strict criteria than
combinatorial equivalence.  Given an appropriate change in criteria, it
will be possible to algorithmically construct a list of 4-conifolds
that are distinct according to that criteria. Of course, the question
that must be addressed is whether or not such a change in criteria is
reasonable.  Obviously, there is no absolute answer to this question.
Such a change in criteria  would not result in a list of
combinatorially inequivalent 4-conifolds, but a list of distinct
4-conifolds that would include more than one instance of the same
combinatorial space. This overcounting of  certain physically distinct
4-conifolds would lead to additional weighting factors in the Euclidean
sum over histories. The effects of such overcounting on  quantum
amplitudes formed in terms of such sums would clearly be an issue for
further study. It is certainly important that such a change in criteria
lead to reasonable results. However, a test of what is reasonable can
only be done by further investigations into the properties of sums over
histories such as \(4simp) constructed with lists of distinct
4-conifolds formed by using different criteria. Therefore a first step
is to present candidates for such alternate criteria.

There are a large number of various possibilities for alternate
criteria for algorithmically classifying 4-conifolds. However, there is
a reasonable requirement to place on such alternate criteria:  The list
of distinct 4-conifolds generated by such criteria should include all
combinatorially inequivalent 4-conifolds, that is the set of physically
distinct 4-conifolds should be a proper subset of this new list.  This
requirement ensures that all classical histories will be included in
the explicit construction of the sum over histories. It also ensures
that the contribution from all possible distinct smooth structures on a
given topological 4-conifold will be included. It follows that all
qualitative results obtained from a semiclassical evaluation of the sum
over histories \(partfn) in terms of Euclidean instantons will be
recovered in a semiclassical evaluation of its concrete implementation
\(fpartfn) in terms of this alternate criteria.  This requirement
therefore provides a good starting point for formulating alternate
criteria that lead to reasonable results.

The possibilities for such alternate criteria can be clearly
illustrated in terms of algorithmic procedures on simplicial complexes;
in addition, such a formulation is  practical
as it will directly lead to implementable sums.  The first procedure
is the most direct;
\proclaim Procedure {(5.1)}. Take the list of all
distinct closed 4-conifolds to be $\cal L$, i.e. that generated by the
procedure for generating an exhaustive list of all closed
4-conifolds.\par
\noindent Note that this method includes all
combinatorially inequivalent
4-conifolds by construction. Of course this set suffers from a massive
amount of redundancy; it includes both 4-conifolds that differ from
each other by a simplicial homeomorphism and 4-conifolds composed of
different numbers of 4-simplices that are combinatorially equivalent.
However, in expressions such as \(4simp), this overcounting of
combinatorially
equivalent 4-conifolds will not result in any manifest divergence as
the same overcounting will occur in both numerator and denominator.
Therefore the main effect of this overcounting is to induce a
particular weighting of physically distinct 4-conifolds; those that are
easiest to build according to the algorithm will be weighted more
heavily than those that are not.

Of course Procedure (5.1) is not aesthetically very nice as it makes no
attempt to eliminate even obvious redundancies. A refinement of this
procedure is the following
\proclaim Procedure {(5.2)}. Beginning with the list $\cal L$ of all
4-conifolds, derive a new list ${\cal L}^1$  in the following
way: For each n, define two 4-conifolds  formed of n simplices
to be equivalent if
they are equivalent under a simplicial homeomorphism. Then
${\cal L}^1$ is the set of all 4-conifolds in $\cal L$ under this
equivalence relation. \par
\noindent This procedure is computable as the number of permutations of
vertices is computable, though large, for each number n.
Note that this set will still include representatives for all
physically distinct
4-conifolds as equivalence under simplicial homeomorphism is a
special case of combinatorial equivalence.  It also clearly
considerably reduces the redundancy present in ${\cal L}$. However,
the list ${\cal L}^1$ will still include redundancies corresponding to
4-conifolds built of different numbers of 4-simplices that are
combinatorially equivalent. A weighting of physically distinct
4-conifolds is again induced by this overcounting and it is an
interesting question as to how much it differs from that of Procedure
(5.1).

One can also generate procedures that partially eliminate the
redundancy caused by combinatorial equivalence. Note that the problem
with determining the combinatorial equivalence or inequivalence of two
spaces lies in the fact that subdivisions of both spaces to arbitrarily
large numbers of simplices are allowed. If only a finite number of
subdivisions are allowed, the number of steps is finite. This
observation can be used to form a finite procedure for {\it weak
combinatorial inequivalence}:
\proclaim Procedure {(5.3)}. Begin with the list ${\cal L}^1$ generated
by procedure 1. Form a sequence ${\cal L}^2_n$ by the following steps:
Begin with the 4-conifold in ${\cal L}^1$ containing the smallest
number of 4-simplices, six, and place it in  ${\cal L}^2_6$. From it
form the set ${\cal S}_7$ of all subdivisions of the element in ${\cal
L}^2_6$ that contain exactly seven 4-simplices. Next find the set
${\cal Q}_7$ of all elements in  ${\cal L}^1$  containing  seven
4-simplices that are not simplicially homeomorphic to any element in
${\cal S}_7$. Add ${\cal Q}_7$
 to ${\cal L}^2_6$ to form  ${\cal L}^2_7$. In general, given ${\cal
L}^2_n$, form the set ${\cal S}_{n+1}$ of all subdivisions of all
elements in ${\cal L}^2_n$ with exactly (n+1) 4-simplices. Next find
the set ${\cal Q}_{n+1}$ all elements in ${\cal L}^1$ that contain
exactly (n+1) 4-simplices that are not simplicially homeomorphic to any
element in ${\cal S}_{n+1}$. Add ${\cal Q}_{n+1}$ to ${\cal L}^2_n$ to
form ${\cal L}^2_{n+1}$. Finally repeat for all n to find ${\cal
L}^2_\infty = {\cal L}^2$.\par

\noindent This procedure eliminates certain combinatorially equivalent
4-conifolds, but not all of them; for example two 4-conifolds
containing different numbers of simplices may not be equivalent if only
subdivisions of the one with fewer simplices  into the one with more
simplices are allowed, but may indeed be equivalent if one allows both
to be further subdivided.  Thus Procedure (5.3) again induces a
weighting on physically distinct 4-conifolds.

These three procedures all provide algorithmically decidable lists of
distinct 4-conifolds. Therefore one can precisely formulate \(4simp)
for 4-conifolds for any of these alternate criteria by taking ${\cal
L}$ to be any of the lists generated by Procedures (5.1) though (5.3).
Thus \(4simp) formulated as a sum over 4-conifolds classified with
respect to alternate criteria provides a concrete computable starting
point for further study of the consequences of topology and topology
change.

Note that the explicit formulation of sums over topological spaces can
be readily extended by application of results in section 4 to any set
of algorithmically decidable spaces. For example, in two dimensions,
the set of 2-pseudomanifolds itself is also algorithmically decidable
and classifiable under combinatorial equivalence. Thus, a  procedure
analogous to that used to construct a list of physically distinct
2-manifolds can be developed that algorithmically constructs a list of
physically distinct 2-pseudomanifolds. Therefore an explicit comparison
of the consequences of a sum over 2-manifolds and a sum over
2-pseudomanifolds can be made.  Hartle carried out a qualitative
analysis of this case and concluded that a sum over 2-pseudomanifolds
resulted in the same qualitative results in the classical
limit,\refto{2dcase} but did not carry out a more explicit computation.
It is clear that the results of such a calculation would be
interesting.\refto{numerical} The case of three dimensions is
particularly interesting as both 3-manifolds and 3-conifolds are
algorithmically decidable; whether or not they are classifiable under
combinatorial equivalence is an open issue for both. However, explicit
implementations of \(4simp) can be formed for both sets of spaces  by
using any of the  Procedures (5.1) through (5.3) described above.
Consequently, three dimensional sums over histories provide an arena in
which the consequences of a sum over 3-conifolds can be tested directly
against results obtained for a sum over 3-manifolds.
Moreover, the issue of the effects of various alternate criteria for
classifying these three dimensional spaces can also be isolated from
that of their topology.\refto{numerical2} In five or six dimensions,
there are no algorithms for generating either n-manifolds or
n-conifolds, but lists of n-pseudomanifolds can be constructed and the
consequences of sums over these spaces can be studied.

Finally, it should  be stressed that the topological issues illustrated
in this section in terms of Regge calculus also apply to sums over
histories not expressly formulated in terms of  simplicial complexes.
By section 3, any combinatorial n-manifold or n-conifold in less than
seven dimensions uniquely corresponds to a smooth n-manifold or
n-conifold respectively; the combinatorial space determines the
topology and smooth charts for the definition of  the smooth space.
Given this smooth space, $<A>_{_{K^n}}$ can be computed in terms of a
functional integral over the space of metrics. It is clear that
changing the calculation of this expectation value in this manner in
\(4simp) in no way changes the properties of the sum over topological
spaces. Thus, the conclusions about the properties of a sum over
topological spaces apply in general.

\head{6.~Conclusion}

In two dimensions, a concrete implementation of a sum over histories
formulation of quantum amplitudes can be carried out explicitly for
2-manifolds.  In four dimensions, such an implementation cannot be made
for manifolds for two reasons; there is no known method of
algorithmically recognizing a 4-manifold and it has been proven that
4-manifolds are not classifiable.  However, 4-conifolds can be
algorithmically described by a simple algorithm and by changing the
criteria for distinctness, explicit algorithms for a sum over
4-conifolds can be formulated. Thus the set of 4-conifolds allows for a
study of the consequences of sums over topology.

Of immediate interest is the question of whether or not enlarging the
set of spaces to be summed over in an expression such as \(4simp)
changes the qualitative results in any unexpected way.
Hartle studied this issue qualitatively in two dimensions and argued
that 2-manifolds would dominate; however, even a qualitative assessment
of the properties of a sum over histories is not as simple in four
dimensions as the action \(regcurv) is no longer topological. Therefore
it is necessary to estimate or evaluate the contribution from
\hbox{$<A>_{_{X^4}}$} for each 4-conifold $X^4$ in order to proceed.
Moreover, one needs some sort of useful and relevant quantity $A$ to
compute when comparing the effects of different choices of algorithms.
The choice of such a quantity is not trivial; generally speaking, one
would like to compute the expectation value of a physical quantity,
that is one that is diffeomorphism invariant. However, known quantities
with this property are topological invariants and thus are not
sensitive to the metric information. Indeed, it is difficult to
formulate explicitly diffeomorphism invariant  quantities that do not
correspond to such invariants. Now, the expectation value of  such
topological invariants may indeed be of interest; however, much of the
interest in explicitly computing a sum such as \(4simp) is in precisely
the consequences of topology on  geometrical quantities.

These problems are not trivial to solve; however, they also do not
present insurmountable obstacles. Different methods of calculating or
estimating $<A>_{_{X^4}}$ can be tried out and expectation values of
different quantities $A$ can be calculated. The results of such trials
will provide information not only relevant to the topological issues
but to questions about the quantum mechanics of gravitational theories
in general.
In any case, it is clear that  explicit formulations of the sum over
histories as found in this paper are such invaluable in any further
investigations of topology and topology change.

\subhead{\undertext{Acknowledgments}}

This work is the outgrowth of results presented by the authors at the
Fifth Marcel Grossman meeting in Perth, Australia, 1988 and elsewhere.
The authors would like to thank the relativity group at the University
of Maryland for their hospitality during the time this work was
initiated.  This work was supported in part by the Natural Science and
Engineering Research Council of Canada, the NSERC International
Fellowship Program and CITA.

\head{ Appendix A: Proof of Lemma (3.3) }

In order to prove Lemma (3.3) it is useful to have the abstract
definition of a simplicial complex. This definition is equivalent to
Def.(2.5) for finite dimensional complexes; however, it does not rely
on the explicit embedding of the complex in Euclidean space. Its
utility is that it can be applied directly in the proof of the Lemma.
\proclaim Definition {(A.1)}. An abstract simplicial complex is a
topological space $|K|$ and a collection of simplices $K$ such
that\hfill\break
$\ \ $ ${i)}$ The set of vertices $K^0$ form a countable set
\hfill\break
$\ \ $ ${ii)}$ The simplices $K$ are a family of subsets of
$K^0$ \hfill\break
$\ \ $ ${iii)}$ Each vertex is contained in only a finite number of
simplices \hfill\break
$\ \  $ ${iv)}$ A set $F\subseteq |K|$ is open if and only if $F\cap
|\sigma|$ is open for all simplices $\sigma \in K$ \hfill\break
$\ \ $ ${v)}$ If a simplex $\tau\subseteq \sigma $ where
$\sigma\subseteq K$ then $\tau \subseteq K$ \hfill\break
$\ \ $ $ {vi)}$ If $\tau, \sigma \in K$, then $\tau \cap \sigma \in K$.
\par

\noindent If the maximum number of vertices contained in any simplex is
less than a fixed number $(n+1)$, then K is finite dimensional and $n$
is its dimension.

Often in the literature, simplicial complexes which satisfy all of the
above conditions are called countable locally finite simplicial
complexes.  Complexes which do not satisfy conditions ${i)}$ and
${ii)}$ are sometimes encountered; however, if a simplicial complex has
finite dimension and is metrizable, then it must satisfy all of the
above conditions. Thus there is no loss of generality in the finite
dimensional case by imposing all of the above conditions on abstract
simplicial complexes.

One can define abstract polyhedra to be the underlying topological
spaces of simplicial complexes. All of the other definitions related to
simplices and polyhedra used in this paper can also be expressed in
abstract terms. For example, simplicial maps are defined to be
continuous maps of vertices to vertices such that the simplices are
also mapped to simplices.  Similarly, a PL map between polyhedra
$f:|K_1|\rightarrow |K_2|$ is a continuous map such that there are
subdivisions of the simplicial complexes $K_1$ and $K_2$ for which the
map $f$ is simplicial.

A characterization  of the relation of Def.(A.1) to Def.(2.5) is given
by the following:
\proclaim Theorem {(A.2)}. Any finite dimensional abstract
simplicial complex $K^n$  embeds in Euclidean space.\par

\noindent First some necessary background: Define ${\bf R}^\infty$ to
be the infinite dimensional vector space consisting of vectors of the
form $(x_1,x_2,\dots,x_k,\dots)$ where each component is real and for
any vector, all but a finite number of its components are nonzero. A
topology is defined on ${\bf R}^\infty$ by the componentwise
convergence of sequences.  Observe that an inner product can be defined
by
$$(x,y)=\sum_{i=1}^\infty x_iy_i$$
\noindent where $x,y\in {\bf R}^\infty$. Although not necessary for the
discussion here, note that the completion of ${\bf R}^\infty$ with
respect to this inner product is the separable Hilbert space
$\ell^2({\bf R})$. Since ${\bf R}^\infty$ is a vector space, simplices
can be defined as the convex hull of affinely independent
points.\refto{points} Thus simplicial complexes in ${\bf R}^\infty$ are
collections of simplices which obey the above abstract definition. One
example of a simplicial complex in ${\bf R}^\infty$ is the single
simplex $\sigma ^\infty$ defined to be the convex hull of a set
orthonormal basis vectors in ${\bf R}^\infty$.  However, simplicial
complexes in ${\bf R}^\infty$ need not be infinite dimensional.
Finally, note that an equivalent definition of a simplicial map between
simplicial complexes in ${\bf R}^\infty$ is that it maps vertices to
vertices and every simplex is mapped linearly to a simplex.

Given any abstract simplicial complex $K$, embed its vertices in the
set of vertices of $\sigma ^\infty$. This embedding of the vertices is
a linearly independent set because it is a subset of the vertices of
$\sigma ^\infty$.  The embedding can be extended to be a simplicial map
\hbox{$f: K\rightarrow \sigma ^\infty$} because the vertices are
embedded in linearly independent way and they completely determine all
of the simplices. For example, a simplex of the form $v_1v_2v_3\dots
v_k$ is mapped to $f(v_1v_2v_3\dots v_k)$ which is equal to
$f(v_1)f(v_2)f(v_3)\dots f(v_k)$ in $\sigma ^\infty$.  One can verify
using the abstract definition of a simplicial complex and simplicial
maps that all simplices in $K$ are mapped to simplices in $\sigma
^\infty$. Moreover, the definition of the topology implies that $f:
|K|\rightarrow |\sigma ^\infty|$ is continuous and that $f(|K|)$ is a
closed subset of ${\bf R}^\infty$. Furthermore, $f(K)$ is a subcomplex
of ${\bf R}^\infty$.

Finally, if $f(K)$ is not a subset of a finite dimensional Euclidean
space, then there is an infinite sequence of simplicial complexes
$f(K)\cap {\bf R}^{m(k)}$ which are strictly increasing sets where
$m(k)$ is map between positive integers such that $m(k_1)<m(k_2)$
whenever $k_1<k_2$.  Hence, either there are  simplices of arbitrary
dimension in $K$, or there are vertices which are in  an infinite
number of simplices.  This contradicts either the fact that $K$ is
finite dimensional or that it is locally finite. Therefore, $f(K)$
embeds in a finite dimensional Euclidean space. Q.E.D.

Given these tools, the proof of Lemma (3.3) follows immediately.  Since
each $S_i$ and $P_i$ is a polyhedron, they all have triangulations.
Furthermore, there are triangulations of $S_1\subseteq P_1$ and
 $S_2\subseteq P_2$ so that the map $\psi$ is simplicial as it is a PL
homeomorphism. Denote such triangulations by $K(S_1)\subseteq K(P_1)$
and $K(S_2)\subseteq K(P_2)$ and the simplicial map corresponding to
$\psi$ by $K(\psi):K(S_1) \to K(S_2)$ where $\psi = |K(\psi)|$. Using
the abstract definition of a simplicial complex it follows that
$$K(P_1\cup_\psi P_2) = K(P_1) \cup_{K(\psi)} K(P_2)$$ is a simplicial
complex where the simplices of $K(S_1)$ and $K(S_2)$ are identified via
$K(\psi)$. Finally, $K(P_1\cup_\psi P_2)$ is finite dimensional if and
only if $P_1$ and $P_2$ are finite dimensional. Therefore
$K(P_1\cup_\psi P_2)$ is a subset of Euclidean space and $P_1\cup_\psi
P_2 =|K(P_1\cup_\psi P_2)|$ is a polyhedron.

\head{ Appendix B: $||E8||$}

It is useful to outline the proof that $||E8||$ has no smooth structure
as a general, readable discussion of this result is not readily
accessible. However, a complete list of references will not be
provided; the interested reader should consult the work of
Freedman\refto{freedman} and Donaldson\refto{donaldson} for a detailed
list.
In order to show that the topological manifold $||E8||$ does not admit
a smooth structure, the following background theorem due to Rohlin is
needed. An orientable smooth 4-manifold admits a spin structure if and
only if one can define spinors on the manifold. As Pl manifolds
uniquely correspond to smooth manifolds in four dimensions, a PL
4-manifold admits a spin structure if its smoothing has one.
Then

\proclaim Theorem (Rohlin). Given any closed PL 4-manifold which admits
a spin structure, then the signature is a multiple of 16.

\noindent The proof of this theorem will be outlined for the case of
smooth 4-manifolds; as  all PL manifolds are homeomorphic to smooth
manifolds in four dimensions, the results can be applied directly to PL
manifolds.  There are direct ways to prove the theorem without using
this fact but for the purposes of the present work it is a simpler
approach to outline.

Recall that the signature of a bilinear symmetric form is the number of
positive eigenvalues minus the number of negative eigenvalues. The
signature of any closed smooth 4-manifold $M^4$ is just the signature
given by the bilinear form
$$Q(\alpha ,\beta )=\int_M \alpha \wedge
\beta \ $$
where $\alpha$ and $\beta $ are any closed 2-forms representing the
second cohomology classes \break $H^2(M^4;{\bf R})$.  Equivalently,
this form can be constructed in terms of the homology classes of the
manifold using the
duality between homology and cohomology as real vector spaces. Then
the signature is the oriented intersection between two  2-surfaces in
general position, each corresponding to an element of the second
homology class $H_2(M^4;{\bf R})$.  Namely, the two 2-surfaces in
general position will intersect in a finite number of points. At each
point of intersection, a value of $+1$ or $-1$ can be assigned
depending on the orientation of the intersections. The intersection
form for each pair of 2-dimensional homology classes will be given by
the sum of $\pm 1$ over all points of intersection and added to give
the total intersection value. This equivalent symmetric bilinear form
on the homology   is also  denoted by $Q$.

For smooth closed 4-manifolds with spin structure, the bilinear form
$Q$ is even, i.e.  $Q(\alpha ,\alpha )\equiv 0\bmod 2$. Since the
existence of spin structure implies that the Dirac operator is well
defined, the Atiyah-Singer index theorem implies the index of the Dirac
operator is an integer and equal to one eighth of the signature. In
four dimensions, spinors are constructed from the usual Clifford
algebra but note that this algebra  has the additional structure that
each nonzero element has a multiplicative inverse, namely, that it is
the division algebra of quaternions. Since multiplicative inverses
exist, they can be used as scalar coefficients  for defining vector
spaces. Since the quaternions are a 4-dimensional real vector space, it
follows that quaternionic vector spaces, that is vector spaces with
quaternionic coefficients, are $4k$-dimensional real or
$2k$-dimensional complex vector spaces where $k$ is any positive
integer. Furthermore, the spinors form such a quaternionic vector
space. Also the Dirac operator commutes with multiplication by
quaternionic numbers. Therefore, the kernel or cokernel of the Dirac
operator is always a quaternionic vector space  because multiplication
of any element in the kernel of the operator by a quaternion results in
another element in the kernel. The useful feature about this
observation is that it means that the kernel and cokernel must be an
$2k$ dimensional complex vector space or equivalently a $4k$ real
vector space. Since the index of the Dirac operator is the difference
of dimensions of the kernel and cokernel, it follows that it is always
even in 4-dimensions.  Therefore, the signature is an even multiple of
8 or equivalently a multiple of 16.

It is useful for ease of presentation to assume that the topological
manifold is simply connected.  This condition simplifies the definition
of the signature and spin structure for nonsmooth manifolds: Given a
simply connected closed smooth 4-manifold, a necessary and sufficient
condition for the manifold to admit a spin structure is that $Q$ is an
even form.  Since $Q$ is defined for any closed 4-manifold not just
smooth 4-manifolds, one can extend this theorem to provide the
definition of a simply connected topological 4-manifold with spin
structure: A closed simply connected topological 4-manifold with a spin
structure is  one with
even $Q$. In order to produce a closed simply connected topological
4-manifold which is not PL, one only need produce a spin manifold
which violates the conclusion of Rohlin's theorem. A candidate for the
intersection form of such a manifold is provided by the exceptional Lie
algebra $E8$.

The exceptional Lie algebra $E8$ is a simple 248-dimensional Lie
algebra with rank 8. Recall that the rank of a Lie algebra is the
dimension of maximal nilpotent subalgebra commonly called the Cartan
subalgebra. For semi-simple Lie algebras the Cartan subalgebra is the
maximal abelian subalgebra. Since the algebra is simple, one can use
the structure constants of the algebra to construct a well defined,
non-singular metric called the Killing metric.  This metric is positive
definite if and only if the Lie group corresponding to the particular
Lie algebra is compact as in the case of $E8$. If the Killing metric is
restricted to the Cartan subalgebra, the matrix representation of this
bilinear form with respect to the root vector basis is

$$\left(\matrix{2&-1&0&0&0&0&0&0\cr
                -1&2&-1&0&0&0&0&0\cr
                 0&-1&2&-1&0&0&0&0\cr
                 0&0&-1&2&-1&0&0&0\cr
                 0&0&0&-1&2&-1&0&-1\cr
                 0&0&0&0&-1&2&-1&0\cr
                 0&0&0&0&0&-1&2&0\cr
                 0&0&0&0&-1&0&0&2\cr}\right).\eqno(biform) $$

The signature of the bilinear form defined by the above matrix is 8.
The Dynkin diagram which is an equivalent graphical representation of
the above matrix and of $E8$ is given in Figure 8a).

So if there is a simply connected 4-manifold with \(biform) as its
intersection form, then it cannot be a PL manifold. Thus it cannot have
a combinatorial triangulation. Such a manifold  indeed exists; it is
$||E8||$.  It can be constructed by the following argument; a more
detailed construction is given at the end of this appendix using the
Dynkin diagram of $E8$.
Let $P=\{ (z,w,u)|z,w,u\in  {\bf C}\  {\rm and }\ z^2+w^3+u^5=1\} $.
This surface is simply connected and has one asymptotic region of the
form $S \times {\bf R}_+ $ where $S=SO(3)/I$ as defined in the example
of the weak triangulation of a 5-sphere given below Def.(2.14).  If the
asymptotic region is cut off at some finite distance so that resulting
4-manifold $E$ has boundary $S$, then the bilinear form $Q$ of the
compact 4-manifold $E$ is \(biform). Furthermore, $E$ is simply
connected by construction. Since $E$ has boundary, one must cap it off
to obtain the desired closed 4-manifold. Freedman proved the following
important and needed result:\refto{freedman}
Given any closed 3-manifold $S$ which is a homology sphere, there
always exists a compact 4-manifold which is contractible to a point and
has boundary $S$.  Since this compact 4-manifold is contractible,
gluing it onto $E$ along the common boundary $S$ will not change the
second homology or the bilinear form of the resulting space. Hence, the
manifold $E$   can be capped off to form a closed 4-manifold commonly
called $||E8||$ which is simply connected and has the same bilinear
form as $E$.  One should note that using $||E8||$ to denote the
4-manifold is an abuse of notation but standard usage.

Immediately, Rohlin's theorem implies $||E8||$ is not a PL manifold as
its signature is not a multiple of 16.  Hence, it does not have a
combinatorial triangulation. Furthermore, using a more careful analysis
it can be shown that it does not have a weak triangulation using
results of Donaldson.\refto{donaldson} Thus $||E8||$ is an example of a
topological manifold that is not homeomorphic to a simplicial complex.
Thus in general there are topological manifolds without
triangulations.

Finally, plumbing techniques will be used to give a more detailed
construction of the manifold $E$.\refto{surgery} The first step is to
observe that the manifolds $S^p\times B^q$ and $S^q\times B^p$ have a
common subspace $B^p \times B^q$ as  $B^p\subset S^p$ and $B^q\subset
S^q$. The plumbing of $S^p\times B^q$ onto $S^q\times B^p$ is then
constructed by taking the disjoint union of $S^p\times B^q$ and
$S^q\times B^p$ and identifying the common subspace $B^p\times B^q$
via the identity map. The resulting manifold $P^{p+q}$ is written
formally as $$P^{p+q}=(S^p\times B^q)\cup_{B^p\times B^q}(S^q\times
B^p)\ . $$
The plumbing of $S^1\times B^1$ onto $S^1 \times B^1$ is illustrated
in Figure 8b).

Although the above plumbing is illustrated using $S^p\times B^q$,
plumbing can also be carried out to form manifolds of dimension $p+q$
using unit disk bundles. Given the tangent bundle of a 2-sphere, the
unit disk bundle is the bundle of all tangent vectors on $S^2$ which
have length less than or equal to one.  The total space of this bundle
has boundary and dimension four.  This bundle is not a trivial product
because there is no nowhere vanishing vector field on the 2-sphere.
However, locally any bundle is a product so locally, the disk bundle
can be written as a trivial bundle. Therefore, if one works on
sufficiently small neighborhoods, the plumbing procedure can be
applied.  Indeed $E$ is constructed by plumbing together unit disk
bundles in the appropriate combination.

In order to construct $E$, start with eight unit disk bundles, one
associated with each circle on the Dynkin diagram of the Lie algebra
$E8$ [see Figure 8a)]. Now, each time a circle on the diagram is
connected with a line, plumb those two spaces.  Let $E$ be the space
resulting from the plumbing procedure. By construction $E$ is a compact
4-manifold with boundary. It can be given a smooth structure. It is
also simply connected because each unit disk space is simply connected
and the gluing is done along simply connected subspaces.

The last step is to verify that this space is $E$ by showing that the
intersection form is the same as the Killing metric restricted to the
Cartan subalgebra. In order to this, one must choose a set of
generators of the homology of $E$. As the unit disk bundle has the same
homology as the 2-sphere, each individual space has a single generator
of the second homology corresponding to that of the $S^2$  before
plumbing. Hence, the homology of $E$ is generated by eight generators,
one for each circle in the Dynkin diagram. If two generators are not
plumbed together, then their intersection form must be zero. This
produces the zero entries of the matrix for $E8$. Now, given two
generators for two unit disk bundles connected by the plumbing, they
can only intersect once. The generators and their intersection for the
plumbing of two $S^1 \times B^1$ is illustrated in Figure 8b).  It is
these generators that produce the off diagonal entries.  Finally, the
intersection of one the generators with itself is given by pushing the
generating sphere a small distance so that it intersects the original
sphere and they are in general position. Basically, this intersects in
two points so it can have a value of 2 or 0. However, for the
two-sphere it is 2 because the intersections do not switch
orientations.  In general, one can repeat the same procedure using the
tangent bundle to show that the intersection
of any manifold with itself is the Euler characteristic which is
consistent with the answer for the 2-sphere.  If one is careful about
signs, one can verify that the intersection form of $E$ is the desired
form \(biform).

\head{ Appendix C: An Undecidable Group}

In Ref.[\cite{boone}], Boone presents a method of constructing
undecidable groups from any Thue system with an unsolvable word
problem. An explicit presentation of such an undecidable finitely
presented group due to Boone is given below. It is based on a Thue
system of Post.\refto{post} The generators of this group are
$$\eqalignno{s_1,s_2,s_3, s_4,q_1,q, t_1,t_2,k,x,y&\cr
l_i, r_i \ \ \ \ \ \ \ \hbox{\rm where}\ \ \ i=1,2,
\ldots,11& \cr}$$
where the notation has been chosen to simplify the presentation of the
relations. The relations are most clearly presented by defining some
auxiliary symbols:

$$\vbox{\settabs 6 \columns
\+ &$\Sigma_1  = s_1q_1$ & & $\Sigma_2 = s_1 q$ & & \cr
\+ &$\Sigma_3  = s_1s_3 $ & &  $\Sigma_4 = s_2 q_1$ & &  \cr
\+ &$\Sigma_5 = s_2q $ &  & $\Sigma_6 = s_2 s_3 $ & & \cr
\+ &$\Sigma_7  = s_3 $ &  & $\Sigma_8 = s_4 s_3 q_1 s_1$ & & \cr
\+ &$\Sigma_9 = s_4s_3qs_2 $ &  & $\Sigma_{10} =
s_4  q_1 s_3$ & & \cr
\+ & & $\Sigma_{11}  = s_4qs_3 $  & & & \cr}$$
and
$$\vbox{\settabs 6 \columns
\+ &$\Gamma_1  = q_1s_1$ & & $\Gamma_2 =  qs_1$ & &  \cr
\+ &$\Gamma_3 = s_3 s_1$ & & $\Gamma_4 =  q_1s_2$ & &  \cr
\+ &$\Gamma_5 =q  s_2 $ & & $\Gamma_6 =s_3  s_2 $ & &  \cr
\+ &$\Gamma_7  = s_3s_4 s_3 $ & & $\Gamma_8 = s_1 q_1 s_4$ & &  \cr
\+ &$\Gamma_9  = s_3qs_4  $ & & $\Gamma_{10} =
s_1  q_1 s_3 s_4$ & & \cr
\+ & &$\Gamma_{11}  = s_2qs_3 s_4 $ & & & \cr}$$
Then the relations of the group are:
$$\vbox{\settabs 6 \columns
\+ & &$ \Sigma_i = l_i\Gamma_i r_i$ & & & \cr
\+ &$ s_\beta l_i = yl_iys_\beta $ & & $  r_i s_\beta =
s_\beta x r_i x $ & &\cr
\+ &$s_\beta y = y y s_\beta  $ & & $  x s_\beta =
s_\beta x x  $ & &\cr
\+ &$t_\alpha l_i = l_i t_\alpha  $ & & $ r_i k =
k r_i $ & & \cr
\+ &$t_\alpha y = y t_\alpha  $ & & $ xk = kx $ & &\cr
\+ & $\ \ \ \ \ \ \ \ \  k q^{-1} t^{-1}_1 t_2 q =
q^{-1} t^{-1}_1 t_2 qk  $ & & & &\cr}$$
where $\alpha = 1,2$ and $\beta = 1,2,3,4$. This finite presentation
has 33 generators and 144 relations; however,  an equivalent
presentation of this group can be derived that has fewer generators and
relations by standard manipulations. However, this particular
presentation is convenient as it can be cleanly written down.
Different undecidable groups can be derived using the Thue systems of
Markov and Scott.\refto{examples} In particular, a finitely presented
group with two generators and 32 relations can be derived from the Thue
system of Scott; however, one of the relations is astronomical in
length.
\references

\refis{excision} The excision property is the following: Let $X$ be a
space with subset
$A$ and $U$ be an open set of $X$ with ${\bar U} \subseteq \hbox{\rm
int} A$. Then $H_*(X,A)= H_*(X-U,A-U)$. See for example,
Ref.[\cite{spanier}].

\refis{points} A set of points $\{ v_i \}$ are affinely independent if
for any fixed vector $v_j$ in the set, the vectors $v_i-v_j$ for $i\neq
j$ are linearly independent for all $v_i$ in the set.

\refis{ST} In particular, one should be aware that modern nomenclature
differs from that of H.~Seifert and W.~Threlfall,
{\it A Textbook of Topology},
 trans. M.~A.~Goldman, (Academic Press, New York, 1980).
 (Original
German edition of {\it Lehrbuch der Topologie} was published in 1934.)
For example, Seifert and Threlfall's definition of a n-manifold is
actually a
 homology manifold in modern terms.

\refis{later} This observation follows from the classification of
2-manifolds and is discussed in more detail in the next section.

\refis{2ndreason} Another reason is that theorems involving invariants
such as the topological signature still hold for PL manifolds but are
not true for topological manifolds admitting only a weak
triangulation.

\refis{3dbreakdown} This proof breaks down in three dimensions because
removing a closed curve changes the fundamental group of a 3-manifold.

\refis{cairns} S.~S.~Cairns, {\sl Bull. Amer. Soc.}
{\bf 41}, 549, (1935).

\refis{boone} W.~W.~Boone, {\sl Ann. Math. }
{\bf 70}, 207, (1959).

\refis{rabin} M.~O.~Rabin, {\sl Ann. Math.}
{\bf 67}, 172, (1958).

\refis{post} E.~L.~Post, {\sl J. Symb. Logic}
{\bf 12}, 1, (1947).

\refis{I} K.~Schleich and D.~M.~Witt, ``{Generalized Sums
over Histories for
Quantum Gravity I: Smooth Conifolds}'', UBC preprint, 1992.

\refis{bigmath} K.~Schleich and D.~M.~Witt, ``Convergence of
Einstein Manifolds to Einstein Conifolds'',  in preparation.

\refis{numerical} K.~Schleich and D.~M.~Witt, ``Summing over
2-manifolds and 2-pseudomanifolds in Regge Calculus'',
in preparation.

\refis{numerical2} K.~Schleich and D.~M.~Witt, in preparation.

\refis{2dcase} J.~B.~Hartle,  {\sl Class.~Quantum~Grav.}
{\bf 2}, 707 (1985)

\refis{jbh1} J.~B.~Hartle, {\sl J.~Math.~Phys.}
{\bf 26}, 804 (1985)
and {\sl J.~Math.~Phys.} {\bf 27}, 287 (1986).
The discussion of Regge techniques in these papers is very useful and
includes extensive references to the literature. However, as detailed
in this section, note that a suggested scheme for constructing all
4-manifolds is flawed and that the classification of simply connected
spin 4-manifolds is misunderstood.

\refis{smoothstructure} E.~E.~Moise, {\sl Ann. Math.}
{\bf 54}, 506, (1951),
{\sl Ann. Math.} {\bf 55}, 172, (1952),
{\sl Ann. Math.} {\bf 55}, 203, (1952),
{\sl Ann. Math.} {\bf 55}, 215, (1952),
{\sl Ann. Math.} {\bf 56}, 96, (1952).

\refis{regge} T.~Regge, {\sl Nuovo Cimento} {\bf 19}, 558 (1961)

\refis{rs} See for example, C.~P.~Rourke and B.~J.~Sanderson,
{\it Introduction to
Piecewise-Linear Topology}, (Springer-Verlag, New York, 1972).

\refis{freedman} See for example, M.~Freedman and F.~Quinn,
{\it Topology of 4-Manifolds},
 (Princeton University Press, Princeton, 1990) and references
therein.

\refis{pIdisc} See for example the discussion of boundary
in section 2 of Ref.[\cite{I}].

\refis{sh} See for example, S.~W.~Hawking, in {\it General
Relativity: An Einstein
Centenary Survey} edited by S. W. Hawking and W. Israel
(Cambridge University Press, Cambridge, 1979).

\refis{genlist} See for example,  J.~B.~Hartle and S.~W.~Hawking,
 {\sl Phys. Rev. D} {\bf 28}, 2960, (1983),
S.~Giddings and A.~Strominger, {\sl Nucl.~Phys. B} {\bf 306},
890 (1988) and  S.~Coleman, {\sl Nucl.~ Phys. B} {\bf 310},
643 (1988), J.~Polchinski, {\sl Nucl. Phys. B} {\bf 325}, 619, (1989).

\refis{greek} C~D.~Papakyriakopoulos, {\sl Bull. Soc. Math. Gr\`ece},
{\bf 22}, 1 (1943)

\refis{spanier} E.~H.~Spanier, {\it Algebraic Topology},
(McGraw-Hill, New York, 1966).

\refis{stillwell} J.~Stillwell, {\it Classical Topology and
Combinatorial
Group Theory}, (Springer-Verlag, New York, 1980).

\refis{toy} R.~Kirby and L.~Siebenmann, {\it Foundational Essays on
Topological Manifolds, Smoothings, and Triangulations},
(Princeton University Press, Princeton, 1977).

\refis{caveat} Strictly speaking, it may be necessary to further
subdivide the simplicial complex with boundary before doubling it over
so that the resulting closed space is also a simplicial complex.
However, any subdivision of the initial complex is combinatorially
equivalent to it by definition and results derived from the subdivided
complex will obviously hold for it as well.

\refis{harsor} J.~B.~Hartle and R.~Sorkin,  {\sl Gen. Rel. Grav.},
{\bf 13}, 6, (1981).

\refis{cheeger} J.~Cheeger, W.~M\"uller and R.~Schrader,
{\sl Comm. Math. Phys.}, {\bf 92}, 405, (1984).

\refis{wordproblem} P.~S.~Novikov, {\sl Trudy. Mat. Inst. Steklov},
{\bf 44}, (1955). See also W.~W.~Boone, "Certain simple,
unsolvable problems
of group theory", V, VI, {\sl Nederl. Akad. Wetensch. Ser. A} {\bf 60}
 22-27, 227-232, (1957).

\refis{examples} D.~Scott, {\sl J. Symb. Logic}, {\bf 21} 111, (1956),
A.~A.~Markov, {\sl Dokl. Akad. Nauk. SSSR} {\bf 55} 587, (1947) and
{\bf 58}, 353 (1947).

\refis{isoproblem} H.~Rogers Jr., {\it Theory of Recursive Functions
and Effective Computibility}, (MIT Press, Cambridge, Massachusetts,
1987),
(Original edition, McGraw-Hill, New York, 1967).

\refis{hamberwilliams} H.~Hamber and R.~Williams, {\sl Nucl. Phys. B}
{\bf 248},
145, (1984).

\refis{4class} A.~A.~Markov, in {\it Proceedings of the International
Congress
of Mathematicians, 1958} (Cambridge Univ. Press, London, 1960).

\refis{surgery} See for example, W.~Browder, {\it Surgery on
Simply-Connected Manifolds}, (Springer-Verlag, New York, 1972).

\refis{kpi1} $K(\pi,n)$ spaces are also known as
 Eilenberg-MacLane spaces. See for example Ref.[\cite{spanier}].

\refis{donaldson} S.~K.~Donaldson and P.~B.~Kronheimer,
{\it The Geometry
of Four-Manifolds}, (Clarendon Press, Oxford, 1990).

\refis{double} J.~W.~Cannon, {\sl Ann. Math.} {\bf 110}, 83, (1979).

\refis{footnote2} Similarly, a compact pure nonbranching simplicial
complex is a n-pseudomanifold if and only if
$H_n(P^n, \partial P^n; {\bf Z}_2 )= {\bf Z}_2$.

\endreferences

\vfill\eject

\head{Figure Captions}

\noindent {\bf Figure 1:}

Two representations, as an illustration and as a list of elements, of a
simplicial complex homeomorphic to a disk are given in a). Another
simplicial complex also homeomorphic to a disk is given in b). The
diagram c)  is not a simplicial complex as  the two distinct edges are
not uniquely specified by the vertices.

\bigskip

\noindent {\bf Figure 2:}

Four examples of 2-dimensional simplicial complexes; a) is not a pure
simplicial complex, b) is not a connected simplicial complex, c) is a
branching simplicial complex and d) is a pure nonbranching complex, but
is not a pseudomanifold.

\bigskip

\noindent {\bf Figure 3:}

Two 2-complexes that are combinatorially equivalent when both are
subdivided.  Both complexes contain the same number of vertices, edges
and faces, but note that all vertices in  a) are contained in four
edges; in b), vertices $a$ and $b$ are in four edges, $c$ and $d$ are
in three edges, and $e$ and $f$ are in five edges.  Thus both must be
subdivided to show combinatorial equivalence.

\bigskip

\noindent {\bf Figure 4:}

A 2-pseudomanifold corresponding to a pinched torus; it is constructed
from the triangulation of the 2-sphere with two disks corresponding to
triangles $bcd$ and $efg$ removed shown in a) by taking the simplicial
cone over the boundary. The star of vertex $a$ consists of the eight
triangles formed from vertices $abcd$ and $aefg$ and all corresponding
subsets. The link of $a$ consists of the edges defined by $abc$ and
$efg$; it is homeomorphic to two disjoint circles.

\bigskip

\noindent {\bf Figure 5:}

The simplicial cone and simplicial suspension of the complex in 1a).
Note that both figures are three dimensional. The cone a) consists of
the two tetrahedra $eabd$ and $ebcd$ and the corresponding faces and
edges. The suspension
 b) consists of four tetrahedra $eabd$, $ebcd$, $fabd$, $fbcd$  and the
 corresponding faces and edges.

\bigskip

\noindent {\bf Figure 6:}

A triangulation of $RP^2$ is given in a); note that the opposite edges
$ab$, $bc$, and $ca$  are actually the same edge. A simplicial cone
and  suspension over $RP^2$ is given in b); again note that opposite
faces $aic$, $cib$, $bia$, $ajc$, $cjb$ and $bja$ are actually the
same. Observe that certain vertex labels and edges have been omitted in
b) for clarity.

\bigskip

\noindent {\bf Figure 7:}

Regge curvature in two and three dimensions; the total curvature is the
sum of the contributions from all n-simplices that contain the
(n-2)-simplex. In a), the contribution to the curvature at $v$ from
$vbc$ is the angle between the two unit vectors $n_1$ and $n_2$ that
lie in edges $vb$ and $vc$. In b), the contribution to the curvature at
$ac$ from $adce$ is the angle between the two unit vectors $n_1$ and
$n_2$ that are orthogonal to $ac$ and lie in faces $aec$ and $adc$.

\bigskip

\noindent {\bf Figure 8:}

The Dynkin diagram of the Lie group E8 is given in a). The plumbing of
$S^1\times B^1$ onto $S^1\times B^1$
  is illustrated in b); the grey disks are identified to form the
linked rings at lower right. The dotted lines are the generators of the
homology and have one point of intersection.

\vfill\eject\vfill\supereject

\end